\documentclass[acmsmall]{acmart}
\acmJournal{CSUR} 

\AtBeginDocument{%
  }

\usepackage{amssymb}

\usepackage{amssymb}
\usepackage{pdflscape}
\usepackage{longtable}
\usepackage{tcolorbox}
\usepackage{longtable}
\usepackage{adjustbox}
\usepackage{colortbl}
\usepackage{balance}
\usepackage{enumitem}
\usepackage{xurl}
\usepackage{makecell}
\usepackage{booktabs}
\usepackage{pdflscape}
\usepackage{graphicx}
\usepackage{rotating} 
\usepackage{pgfplots}
\pgfplotsset{compat=1.18}

\usepackage{calc}  

\usepackage{booktabs}
\usepackage[table]{xcolor}
\usepackage{array}
\usepackage{pifont}  

\usepackage{tikz}

\pgfplotsset{compat=1.18}

\usepackage{cleveref}
\usepackage{hyperref}

\definecolor{darkgreen}{RGB}{0,100,0}
\definecolor{darkblue}{rgb}{0.0, 0.0, 0.5}
\definecolor{lightergray}{rgb}{0.93, 0.93, 0.93}
\definecolor{rowgray}{gray}{0.96}
\definecolor{headergray}{gray}{0.90}
\definecolor{darkgreen}{RGB}{0,120,0}

\usepackage{todonotes}

\usepackage{xcolor}

\newcounter{reviewer}
\newcounter{comment}[reviewer]

\makeatletter
\providecommand{\shorttitle}{Response to Reviewers}
\providecommand{\@acmArticle}{}
\makeatother

\begin{document}

\title[A Systematic Literature Review of LLMs in Code Security]{From Vulnerabilities to Remediation: A Systematic Literature Review of LLMs in Code Security}

\author{Enna Basic}
\affiliation{%
  \institution{Department of Computer Science, Örebro University}
  \city{Örebro}
  \country{Sweden}
}
\affiliation{%
 \institution{Epiroc Rock Drills}
 \city{Örebro}
 \country{Sweden}
 }
 
\author{Alberto Giaretta}
\affiliation{%
 \institution{Department of Computer Science, Örebro University}
 \city{Örebro}
 \country{Sweden}
 }

\begin{abstract}

Large Language Models (LLMs) have emerged as powerful tools for automating programming tasks, including security-related ones. 
However, they can also introduce vulnerabilities during code generation, fail to detect existing vulnerabilities, or report nonexistent ones.
This systematic literature review investigates the security benefits and drawbacks of using LLMs for code-related tasks. In particular, it focuses on the types of vulnerabilities introduced by LLMs when generating code. Moreover, it analyzes the capabilities of LLMs to detect and fix vulnerabilities, and examines how prompting strategies impact these tasks. Finally, it examines how data poisoning attacks impact LLMs performance in the aforementioned tasks.


\end{abstract}

\ccsdesc[500]{Security and privacy~Software security engineering}
\ccsdesc[500]{Security and privacy~Vulnerability scanners}
\ccsdesc[300]{Computing methodologies~Artificial intelligence}

\keywords{Large Language Models, LLMs, Security Vulnerabilities, LLM-generated Code, Vulnerability Detection, Vulnerability Fixing, Prompt Engineering, Data Poisoning}


\maketitle

\section{Introduction}\label{sec:intro}
Large Language Models (LLMs) are Machine Learning (ML) models designed for natural language processing~\cite{blank2023large}.  
Representative models such as ChatGPT~\cite{OpenAI_ChatGPT}, Llama~\cite{touvron2023llama} and GitHub Copilot~\cite{copilot} gained extreme popularity and recognition in the last couple of years due to their ability to perform well across different tasks. Throughout this paper, we use the term ``LLM'' to refer to large autoregressive Transformer language models that support prompt-driven, open-ended code generation and analysis. In contrast, we use Pre-trained Language Model (PLM) to refer to pre-trained (often smaller) encoder-only Transformer models (e.g., CodeBERT, GraphCodeBERT) that are typically fine-tuned for task-specific prediction or classification rather than open-ended generation. We treat these PLMs and other task-specific neural models as traditional ML baselines and include them only when they serve as comparison points for LLM-based approaches.

Among other capabilities, LLMs effectively translate natural language into code, assist in code comprehension, and answer questions related to code~\cite{10.1145/3631802.3631806}. Many developers utilize these abilities to optimize coding processes, reducing development time and thereby improving productivity. According to GitHub statistics, GitHub Copilot is now responsible for generating approximately 46\% of code and boosts developers' coding speed by up to 55\%~\cite{copilotStatistics}.
In addition to their coding skills, LLMs can be powerful tools for vulnerability detection and fixing, particularly when provided with detailed prompts and clear background information~\cite{le2024study, zhou2024large}.


While LLMs represent an advancement in code generation, they exhibit several limitations that pose security risks. One concern is that, despite their proficiency in producing functional code, these models may generate code that violates secure-coding best practices~\cite{pearce2022asleep}.
Another concern is that LLMs are trained on enormous datasets, often gathered from unverified online platforms (e.g., GitHub~\cite{gitHub} and HuggingFace~\cite{huggingFace}). This reliance on potentially insecure data introduces data poisoning risks, where an attacker inserts harmful data samples into the training set~\cite{cotroneo2024vulnerabilities}. These and other shortcomings can lead LLMs to generate insecure code, as well as influence their abilities to detect and fix security vulnerabilities.

Regardless of the risks, LLMs are largely used for their code-related capabilities. Therefore, to take informed decisions, it is crucial for practitioners to have a clear understanding about the precise nature of risks and benefits. In this Systematic Literature Review (SLR), we aim to:
\begin{itemize}
    \item Explore the potential security vulnerabilities introduced by LLM-generated code;
    
    \item Evaluate the effectiveness of LLMs in detecting and fixing vulnerabilities, as well as the prompting strategies used for these tasks;

    \item Investigate the effects of data poisoning on LLMs  ability to produce secure code and to detect and fix vulnerabilities effectively.

\end{itemize}

\subsection{Outline}\label{subsec:outline}

This paper is organized as follows. Section~\ref{sec:relatedWork} describes related works. Section~\ref{sec:methodology} describes the methodology we followed in this work. Section~\ref{sec:indroducedVulnerabilities} discusses the vulnerabilities identified in LLM-generated code, by the state of the art. Section~\ref{sec:detection} examines LLMs ability to detect code vulnerabilities, Section~\ref{sec:fixing} their ability to fix pre-identified vulnerabilities, and Section~\ref{sec:detectionAndFix} their potential to detect and fix vulnerabilities in a unified process. Furthermore, Section~\ref{sec:prompting} covers prompting techniques used for vulnerability detection and fixing, as described by the studies covered in this review.
Section~\ref{sec:poisoning} explores the impact of poisoned training data on LLMs ability to generate secure code, detect vulnerabilities, and fix them. Section~\ref{sec:discussion} addresses the research questions, open challenges, and threats to validity. Finally, Section~\ref{sec:conclusion} concludes the paper.

\section{Related Work}
\label{sec:relatedWork}

In recent years, the literature has focused on different aspects of the intersection between LLMs and code security.
As we discussed in Section~\ref{sec:intro}, in this SLR we focus on three main thematic areas. Although other surveys covered these areas, they all focused on only one (or two) areas at a time. To the best of our knowledge, this is the first comprehensive survey that addresses all three areas.
Furthermore, this is the first paper to attempt to categorize the various vulnerabilities that can be introduced by LLMs. Among the surveys covered in this section, only
Negri et al.~\cite{negri2024systematic} describe how LLMs often generate code containing specific vulnerabilities, but they did not provide a systematic categorization of such vulnerabilities. In contrast, our work organizes them into 10 distinct categories.

Other characteristics differentiate our paper from the current literature. First, while other surveys investigated the performance of LLMs for vulnerability detection and fixing, unified processes for addressing both tasks have not been surveyed. Second, no other survey investigates how the prompting strategy of choice impacts tasks related to code security.
Yao et al.~\cite{YAO2024100211} reviewed various applications of LLMs in security, including vulnerability detection and fixing.  Similarly, Sheng et al.~\cite{sheng2025llms} surveyed LLM-based software vulnerability detection, focusing on detection techniques, datasets, evaluation metrics, and open challenges. 
Taghavi et al.~\cite{taghavi2025large} reviewed models, methods, techniques, datasets, and metrics for LLM-based software vulnerability detection. Germano et al.~\cite{germano2025systematic} reviewed the use of LLMs for vulnerability detection, repair, and explanation in source code, noting that most studies focus on vulnerability detection, with fewer addressing repair and explanation.
However, as shown in Table~\ref{tab:relatedwork-compare}, none of these reviews covers the full scope of our work, namely, secure code generation, unified detection and fixing, the impact of prompting strategies, and the role of data poisoning in code-security tasks.
Although Zhou et al.~\cite{zhou2024largelr} provided an overview of prompting techniques, listing which papers applied each technique for vulnerability detection and repair, their work did not investigate the impact of prompting techniques on such tasks, nor the merging of vulnerability detection and fixing in a unified process.

Furthermore, the literature only partially covers the impact of data poisoning attacks on LLMs code generation, and our work fills this gap by analyzing this potential impact. Yao et al.~\cite{YAO2024100211} introduce the threat of data poisoning attacks, but do not discuss the deeper effects of these attacks on LLMs secure code generation nor on vulnerability-related tasks. 
Similarly, Chen et al.~\cite{chen2024security} highlight the general consequences of poisoning attacks on general code generation tasks, including the effects on code summarization and code search, unrelated to the security focus of this review.
In this paper, we provide an in-depth exploration of security-specific aspects, discussing how poisoning attacks influence the ability of LLMs to generate secure code, detect vulnerabilities, and propose fixes. More broadly, several surveys review LLMs across cybersecurity or software engineering tasks beyond code security~\cite{xu2024large, hou2023large, zhang2025llms, zhang2024llms}, falling outside of the scope of this paper.

Besides these differences in scope and thematic coverage, this SLR is further distinguished by the recency of the literature it synthesizes. 
To the best of our knowledge, our work incorporates all studies published through 2025 and into early 2026, whereas previous studies cover 2025 only partially (also shown in Table~\ref{tab:relatedwork-compare}). In a fast-evolving area such as LLMs in code security, this more up-to-date coverage strengthens the relevance and timeliness of our synthesis.

\newcommand{\ccell}[1]{\parbox[c]{\linewidth}{\centering #1}}

\begin{table*}[tb]
    \renewcommand{\arraystretch}{1.2}
    \footnotesize
    \centering
    \caption{Comparison of prior secondary studies on LLMs and code security. Each column indicates whether a review explicitly addresses the corresponding topic.}
    \label{tab:relatedwork-compare}

    \setlength{\tabcolsep}{4pt}

    \begin{tabular}{
        p{0.155\linewidth}
        p{0.105\linewidth}
        p{0.15\linewidth}
        p{0.07\linewidth}
        p{0.07\linewidth}
        p{0.10\linewidth}
        p{0.10\linewidth}
        p{0.09\linewidth}
    }
        \toprule
        \rowcolor{white}
        \textbf{Secondary study} &
        \ccell{\textbf{Month/Year}} &
        \ccell{\textbf{Vuln.\ categorization}} &
        \ccell{\textbf{Detect}} &
        \ccell{\textbf{Fix}} &
        \ccell{\textbf{Detect+Fix}} &
        \ccell{\textbf{Prompting}} &
        \ccell{\textbf{Poisoning}} \\
        \midrule

        Negri et al.~\cite{negri2024systematic}     & \ccell{04/2024} & \ccell{\checkmark} & \ccell{} & \ccell{} & \ccell{} & \ccell{} & \ccell{} \\
        \rowcolor{lightergray}

        Yao et al.~\cite{YAO2024100211}             & \ccell{01/2024} & \ccell{} & \ccell{\checkmark} & \ccell{\checkmark} & \ccell{} & \ccell{} & \ccell{} \\

        Sheng et al.~\cite{sheng2025llms}           & \ccell{09/2025} & \ccell{} & \ccell{\checkmark} & \ccell{} & \ccell{} & \ccell{} & \ccell{} \\
        \rowcolor{lightergray}

        Taghavi et al.~\cite{taghavi2025large}      & \ccell{01/2025} & \ccell{} & \ccell{\checkmark} & \ccell{} & \ccell{} & \ccell{} & \ccell{} \\

        Germano et al.~\cite{germano2025systematic} & \ccell{11/2025} & \ccell{} & \ccell{\checkmark} & \ccell{\checkmark} & \ccell{} & \ccell{} & \ccell{} \\
        \rowcolor{lightergray}

        Zhou et al.~\cite{zhou2024largelr}          & \ccell{11/2024} & \ccell{} & \ccell{\checkmark} & \ccell{\checkmark} & \ccell{} & \ccell{\checkmark} & \ccell{} \\

        Chen et al.~\cite{chen2024security}         & \ccell{04/2025} & \ccell{} & \ccell{} & \ccell{} & \ccell{} & \ccell{} & \ccell{\checkmark} \\
        \midrule

        \rowcolor{lightergray}
        \textbf{This SLR} & \ccell{01/2026} & \ccell{\checkmark} & \ccell{\checkmark} & \ccell{\checkmark} & \ccell{\checkmark} & \ccell{\checkmark} & \ccell{\checkmark} \\
        \bottomrule
    \end{tabular}
\end{table*}

\raggedbottom

\section{Methodology}
\label{sec:methodology}
 
In this study, we conducted an SLR following established guidelines proposed by Petersen et al.~\cite{PETERSEN20151}.Following the adopted guidelines, we carried out the review through three structured phases: planning, conducting, and reporting. In this section, we present the research questions, search strategy, selection criteria, study selection process, and the publication trends and venue distribution of the included studies.

\subsection{Research Questions}
\label{subsec:RQs}

We formulated the following Research Questions (RQs) to guide our investigation and help fulfill the aims of this study:

 \begin{itemize}
    \item \textbf{RQ1:} What security vulnerabilities in LLM-generated code have been reported in the current literature?
    
    \item \textbf{RQ2:} What is the state of the art on the capabilities and limitations of LLMs for detecting and fixing security vulnerabilities in code?
       
    \begin{itemize}
    \item \textbf{RQ2.1:} Which prompting techniques are used in LLM-based detection and fixing of security vulnerabilities in code, and how do these techniques affect model performance?
  
    \end{itemize}

    \item \textbf{RQ3:} What effects of poisoned training data on LLMs ability to generate secure code, detect vulnerabilities, and fix them have been reported in the current literature?

\end{itemize}

\subsection{Search Strategy}
\label{subsec:searchStrategy}

To address the RQs, we developed the following base search strings for the automated database search. We adapted them to accommodate database-specific syntax and query constraints while preserving their logical structure. The first search string addresses RQ1:

\begin{tcolorbox}[colframe=black, colback=lightergray]
\footnotesize
(``Large Language Models” OR ``Language Model” OR ``LLMs” OR CodeX OR Llama OR Copilot OR GPT-* OR ChatGPT)
AND
(``Security Vulnerabilities” OR ``Security Risks” OR ``Security Flaws” OR ``Software Security” OR ``Impact On Code Security” OR Cybersecurity OR Vulnerabilities)
AND
(``Code Generation” OR ``AI-generated Code” OR ``Automated Code Generation”)
\end{tcolorbox}

The second search string is designed to address RQ2:
\begin{tcolorbox}[colframe=black, colback=lightergray]
\footnotesize
(``Large Language Models” OR ``Language Model” OR CodeX OR Llama OR Copilot OR GPT-* OR ChatGPT) AND
(``Vulnerability Detection” OR ``Bug Detection” OR ``Security Flaw Detection” OR ``Code Analysis” OR ``Static Analysis”
 OR ``Vulnerability Remediation” OR ``Bug Fixing” OR ``Automated Code Repair” OR ``Security Patch” OR ``Code Patching”)   AND
(Prompt* OR ``Prompt Engineering” OR ``Prompting Strategy” OR ``Prompt-based” 
 OR ``Zero-shot” OR ``Few-shot” OR ``One-shot” OR ``Chain-of-Thought” OR ``In-context”)  

\end{tcolorbox}

Finally, the third search string is designed to address RQ3:
\begin{tcolorbox}[colframe=black, colback=lightergray]
\footnotesize
(``Large Language Models” OR ``Language Model” OR LLMs OR CodeX OR Llama OR Copilot OR GPT-* OR ChatGPT) 
AND (``Code Generation” OR ``AI-generated Code” OR ``Automated Code Generation”) 
AND (``Security Vulnerabilities” OR ``Security Risks” OR ``Security Flaws” OR ``Software Security” OR ``Impact On Code Security” OR ``Cybersecurity” OR Vulnerabilities) 
AND (``Training Data Poisoning” OR ``Poisoned Datasets” OR ``Data Poisoning Attacks” OR ``Adversarial Attacks” OR ``Malicious Training Data”) 
AND (``Vulnerability Detection” OR ``Bug Detection” OR ``Security Flaw Detection” OR ``Code Analysis” OR ``Static Analysis” OR ``Vulnerability Remediation” OR ``Bug Fixing” OR ``Automated Code Repair” OR ``Security Patch” OR ``Code Patching”)

\end{tcolorbox}

\subsection{Selection Criteria}
\label{subsec:criteria}

To ensure a relevant review of the literature, we defined inclusion and exclusion criteria.  The detailed criteria for paper selection are outlined in Table~\ref{tab:inclusionExclusionCriteria}. 
Given that the field is relatively new and has seen significant contributions in the last few years, we did not establish criteria for narrowing down the period of publications. We applied these inclusion and exclusion criteria throughout the study selection process, which is described in the following subsection.


\renewcommand\arraystretch{1.2}
\begin{table}[tb]
\centering
\footnotesize 
\caption{Inclusion and exclusion criteria for paper selection.}
\label{tab:inclusionExclusionCriteria}
\begin{tabular}{p{13.5cm}}
\rowcolor{lightergray}
\textbf{Inclusion Criteria} \\ \toprule

\textbf{IN1:} Peer-reviewed full-text papers identified through the database search and snowballing. \\
\textbf{IN2:} Papers that specifically identify vulnerabilities in code generated by LLMs. \\
\textbf{IN3:} Papers that evaluate the capability of LLMs to detect and fix security vulnerabilities in code.\\
\textbf{IN4:} 
Papers that explore the impact of poisoned training data on LLMs ability to generate secure code and detect and fix vulnerabilities. \\

\bottomrule
\textbf{Note:} To be included in the review, a paper must satisfy \textbf{IN1} \textbf{and} \textbf{(IN2 \textbf{or} IN3 \textbf{or} IN4)}.\\ 
\bottomrule

\addlinespace[4pt]

\rowcolor{lightergray}
\textbf{Exclusion Criteria }\\
\toprule
\textbf{EX1:} Papers not written in English.\\
\textbf{EX2:} Articles without accessible full-text versions. \\
\textbf{EX3:} Secondary studies, such as SLRs or surveys.\\
\textbf{EX4:} Papers that focus on LLM capabilities unrelated to security. \\
\textbf{EX5:} Studies evaluating only PLMs without involving LLMs, and studies in which the main proposed approach is based on PLMs rather than LLMs. \\
\textbf{EX6:} Studies centered on agent-based LLM systems. \\
\textbf{EX7:} Papers focusing exclusively on smart contract vulnerabilities.\\

\bottomrule

\end{tabular}
\end{table}
\raggedbottom


\subsection{Study Selection Process}
\label{subsec:studySelection}

\begin{figure*}[tb]
    \centering
    \includegraphics[width=1\textwidth]{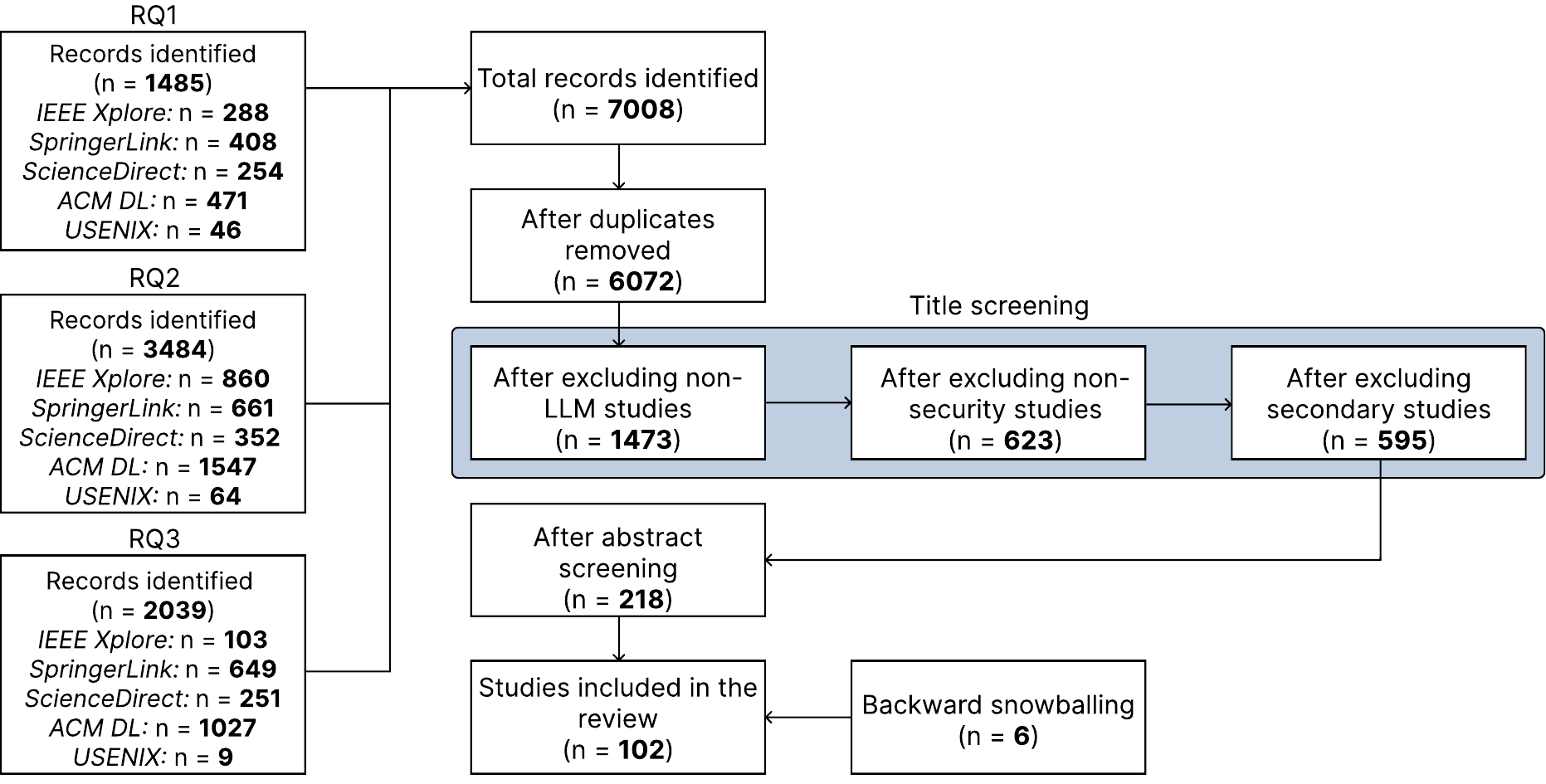}  
    \captionsetup{justification=centering}
    \caption{PRISMA-based flow diagram of the study identification and screening process.}
    \label{fig:methodology}
\end{figure*}

We conducted an automated search across IEEE Xplore, ACM Digital Library, ScienceDirect, SpringerLink, and USENIX database. The search was executed between January 7 and January 12, 2026. The study identification and screening procedure is reported using a PRISMA-based flow diagram~\cite{page2021prisma}, as shown in Figure \ref{fig:methodology}.

We retrieved 7008 records from the automated search. After removing duplicates, we obtained 6072 unique records. We then conducted title screening in three steps. First, we excluded papers not related to LLMs, leaving 1473 records. Second, we excluded papers not related to security, reducing the set to 623 records. Third, we excluded secondary studies (e.g., surveys and SLRs), resulting in 595 papers that proceeded to abstract screening. We then screened abstracts and selected 218 studies for full-text assessment. After full-text review, we included 96 primary studies. Finally, backward snowballing identified 6 additional relevant papers, resulting in 102 included studies.

\subsection{Publication Trends and Venue Distribution}
\label{subsec:trendsAndVenues}

Given the fast-evolving nature of this research area, we provide an overview of the included studies by examining their distribution over time and across publication venues.

\begin{figure}[tb]
    \centering
    \begin{minipage}{0.49\linewidth}
        \centering
        \includegraphics[width=\linewidth, trim=4mm 3mm 6mm 4mm, clip]{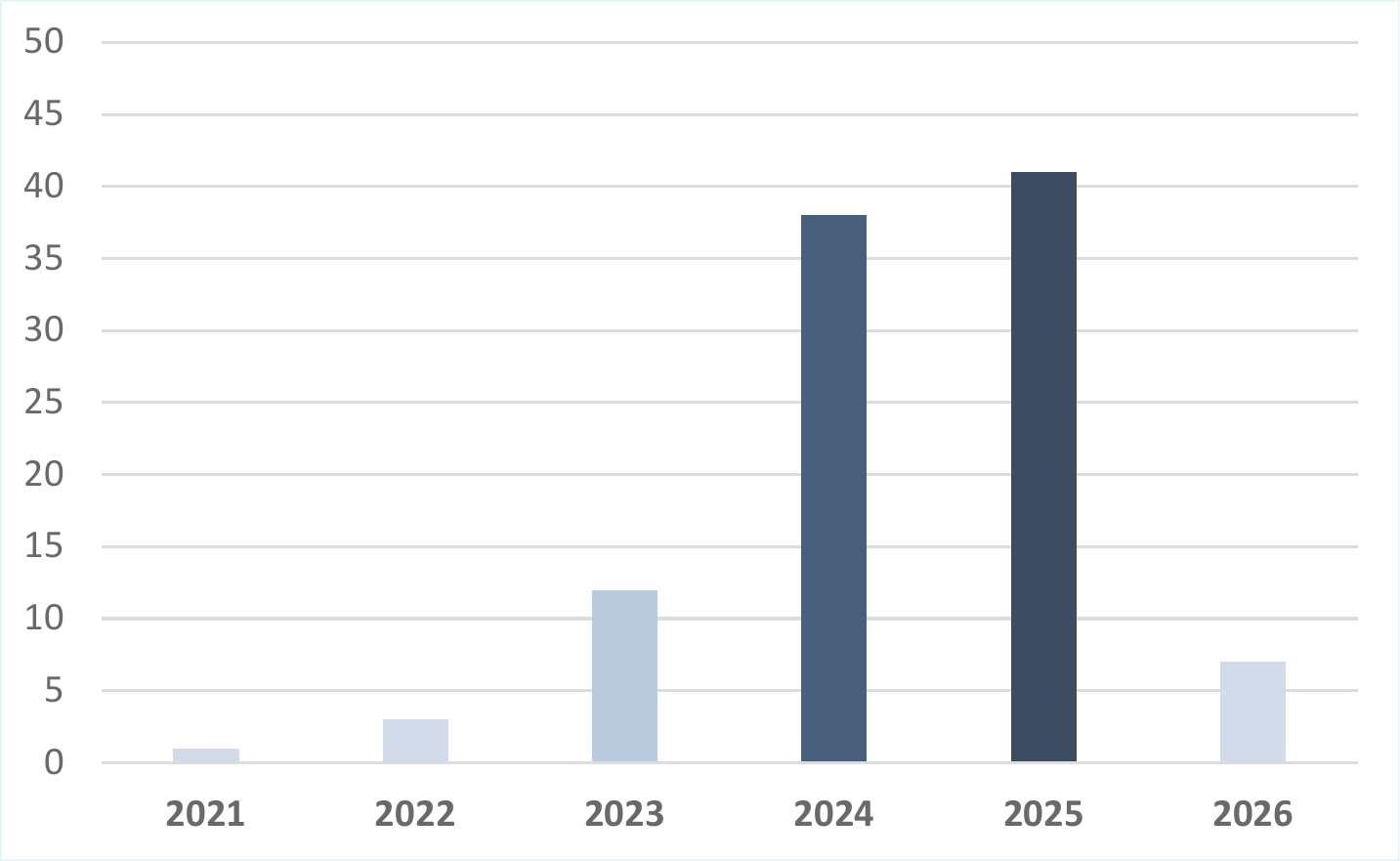}
        \caption{Distribution of studies by publication year.}
        \label{fig:years}
    \end{minipage}%
    \hfill
    \begin{minipage}{0.49\linewidth}
        \centering
        \includegraphics[width=\linewidth, trim=4mm 3mm 6mm 4mm, clip]{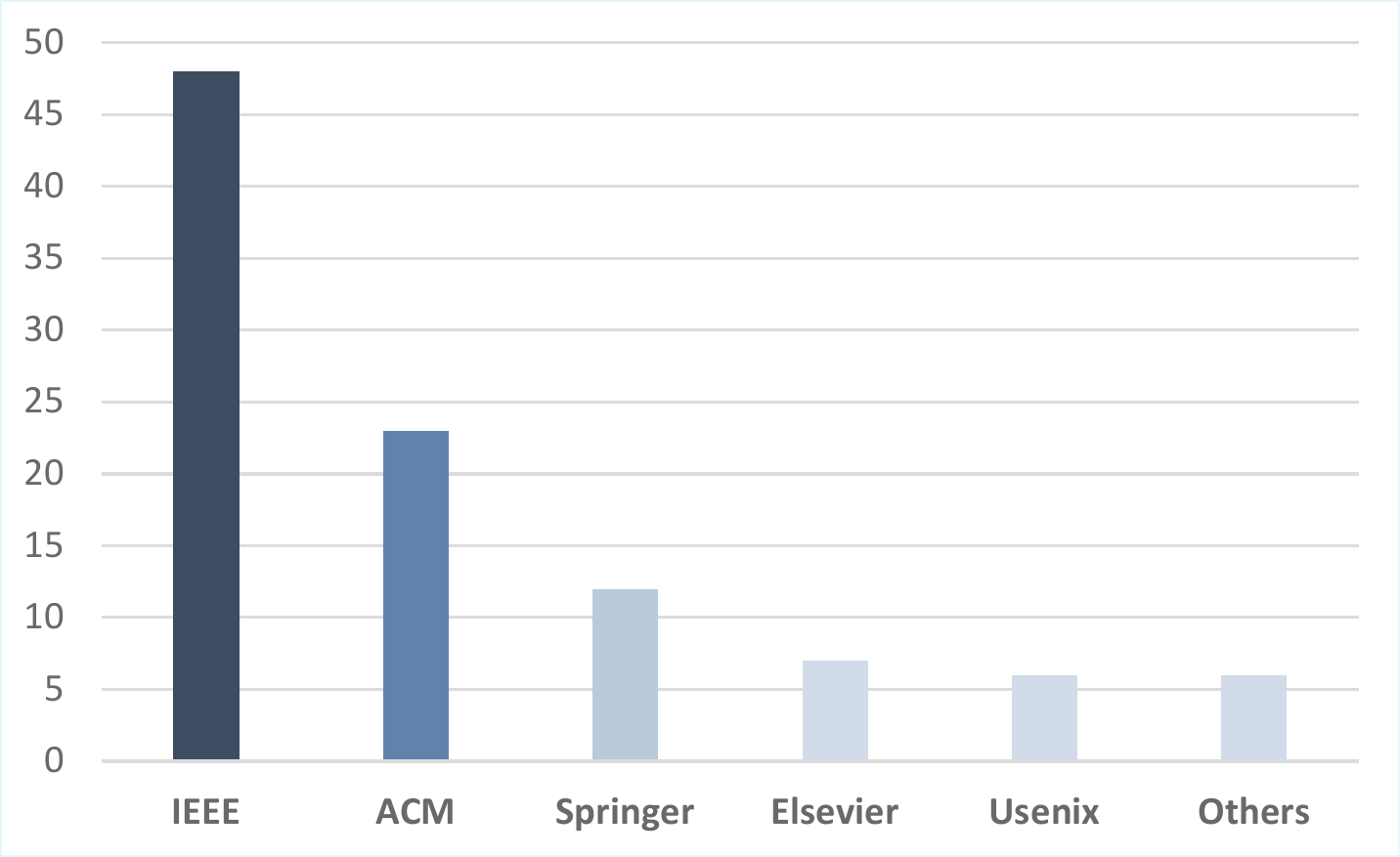}
        \caption{Distribution of studies across venues.}
        \label{fig:venues}
    \end{minipage}
\end{figure}

As shown in Figure~\ref{fig:years}, the number of publications has increased in recent years. Only 4 studies were published between 2021 and 2022 (one in 2021 and three in 2022), followed by 12 studies in 2023. The number of publications increased further in 2024 and 2025, with 38 and 41 studies, respectively. The lower number observed for 2026 (7 studies) is expected because the search was conducted in January 2026 and therefore does not represent a complete publication year. 

Figure~\ref{fig:venues} shows the distribution of studies across publication venues. IEEE venues account for the largest share of the included studies (48 papers), followed by ACM (23) and Springer (12). Elsevier, USENIX, and other venues contribute smaller numbers of studies (7, 6, and 6 papers, respectively). 
Overall, most of the included studies were published in IEEE and ACM venues.

\section{Security Vulnerabilities Introduced by LLMs}
\label{sec:indroducedVulnerabilities}

Although LLMs became popular for their ability to generate functional code fast and speed up development, they can introduce security vulnerabilities that developers might miss, leading to risks regarding code integrity and safety~\cite{he2023large}. 
In this section, we address RQ1 by focusing on specific vulnerabilities introduced by LLM-generated code, as identified in 21 studies, and on how these map to the Common Weakness Enumeration (CWE) taxonomy~\cite{cwe}. For each study, we extracted the reported vulnerabilities and grouped them into 10 categories that we defined for this SLR, which are summarized in Table~\ref{tab:vulnerabilitiesCheckmarks}. In each cell, we either list the CWE(s) reported in the study for that category or use a green checkmark (\textcolor{darkgreen}{\checkmark}) to indicate that the study reported vulnerabilities belonging to that category without specifying the corresponding CWE identifiers.

We do not compare vulnerability rates across studies because the primary studies rely on different code-generation tasks, datasets, and evaluation setups, which makes rate-based comparisons inconsistent.
We summarize the models, programming languages, and datasets used in each study in Appendix~\ref{app:appendix} (Table~\ref{tab:rq1_models_languages_datasets}).

\begin{table}[tb]
  \centering

  \renewcommand\arraystretch{1.35}  
  \setlength{\tabcolsep}{3.2pt}     
  \footnotesize 
                    
  \caption{Vulnerability categories in LLM-generated code across the reviewed studies. Cells list CWE ID(s) when reported. Green checkmarks (\textcolor{darkgreen}{\checkmark}) indicate the category was reported without specific CWE(s).}
  \label{tab:vulnerabilitiesCheckmarks}

  \rowcolors{3}{lightergray}{white}

  \begin{adjustbox}{width=\textwidth}
    \begin{tabular}{lcccccccccc}
      \toprule
    
      \rotatebox{90}{References} &
      \rotatebox{90}{Injection} &
      \rotatebox{90}{\makecell{Memory\\Management}} &
      \rotatebox{90}{\makecell{File\\Management}} &
      \rotatebox{90}{Deserialization} &
      \rotatebox{90}{\makecell{Sensitive Data\\Exposure}} &
      \rotatebox{90}{\makecell{Authentication\\and\\Authorization}} &
      \rotatebox{90}{Cryptography} &
      \rotatebox{90}{\makecell{Resource\\Management}} &
      \rotatebox{90}{\makecell{Coding\\Standards}} &
      \rotatebox{90}{\makecell{Error\\Handling}} \\
      \midrule

      \cite{liu2024no}
      & 78, 79, 89 & \makecell{119, 125, 190,\\ 416, 476, 787} & 22, 434 & 502 & 200, 206 & 522, 798 & & & 20 & \\

      ~\cite{khoury2023secure}
      & 94, 564 & 190 & 35, 377 & 502 & 256 & & 323, 338, 759 & 400 & 133, 453, 687 & \\

      ~\cite{asare2023github}
      & 707 & & & & & 284 & & 664 & 682, 691, 710 & \\

      \cite{perry2023users}
      & \textcolor{darkgreen}{\checkmark} & \textcolor{darkgreen}{\checkmark} & \textcolor{darkgreen}{\checkmark} & & & & \textcolor{darkgreen}{\checkmark} & & & \\

      \cite{sandoval2023lost}
      & & \makecell{401, 416, 476,\\ 787} & & & & & & & 758 & 252 \\

      \cite{pearce2022asleep}
      & 77, 78, 79, 89 & \makecell{119, 125, 190,\\ 416, 476, 787} & 22, 434 & & 200 & 306, 522, 798 & & & 20 & \\

      \cite{siddiq2023generate}
      & 79, 918 & & & & \makecell{208, 209, 215,\\ 312} & 287, 798 & 295, 338 & & & \\

      \cite{hajipour2024codelmsec}
      & 79 & 190, 787 & 22 & & & & & & 20 & \\

      \cite{he2023large}
      & 78, 79, 89 & \makecell{125, 190, 476,\\ 787} & 22 & & & & & & & \\

      \cite{siddiq2022securityeval}
      & 79, 89, 94 & & 22 & 502 & & \makecell{297, 352, 601,\\ 730} & 327 & & 20, 611 & \\

      \cite{toth2024llms}
      & 80, 83 & & & & & & & & 20 & \\

      \cite{tihanyi2025secure}
      & & \makecell{120, 129, 190,\\ 191, 476, 761,\\ 787, 788, 843} & & & & & & & 628 & \\

      \cite{hamer2024just}
      & & & & & & 798 & 327, 328 & 404, 772 & & \\

      \cite{rabbi2024ai}
      & 78, 94 & & & 502 & & 259 & 327, 330 & 400, 605 & 20 & 703 \\

      \cite{fu2023security}
      & \makecell{78, 89, 94} & & 22 & 502 & 215, 312 & 798 & 295 & 367, 772 & \makecell{369, 563, 617,\\ 628} & 117 \\

      \cite{he2024instruction}
      & 79, 89 & 119 & 22 & 502 & 200 & 352, 732, 915 & \makecell{295, 326, 327, 338} &  & 611 & \\

      \cite{gong2024well}
      & \makecell{78, 79, 80,\\ 90, 94, 95,\\ 99, 113, 116,\\ 643, 918, 943}
      & & \makecell{22, 377,\\ 434, 641} & 502
      & \makecell{200, 209, 215,\\ 319, 385}
      & \makecell{250, 259, 269,\\ 283, 285, 425,\\ 522, 601, 732,\\ 798, 841, 941}
      & \makecell{321, 326, 329,\\ 330, 331, 759,\\ 760, 1204}
      & \makecell{367, 400, 406,\\ 414, 605, 776}
      & \makecell{454, 462, 595,\\ 611, 827}
      & 252, 117 \\

      \cite{jamdade2024pilot}
      & 79, 89, 93 & & & & 200 & & & & & \\

      \cite{aydin2025security}
      & \makecell{79, 89, 94,\\ 95, 78, 134,\\ 601, 918}
      & & 22 & & 200, 208, 539 & 319, 384 & 327, 347, 916 & 606, 770
      & \makecell{565, 1188, 1287,\\ 1321, 1333}
      & 117, 614 \\

      \cite{majdinasab2024assessing}
      & 78, 79, 89 & & 22, 434 & 502 & 200 & 306, 522, 732 & & & & \\

      \cite{firouzi2024time}
      &  & &  &   &   & 259 & 327, 798, 330, 326 & & & \\

      \bottomrule
    \end{tabular}
  \end{adjustbox}
\end{table}

\subsection{Injection Vulnerabilities}
Injection vulnerabilities are identified across most of the reviewed studies, appearing in 17 papers.
This category contains various types of vulnerabilities, including SQL injection, Cross-Site Scripting (XSS), OS command injection, and regular expression (regex) injection. 
Two studies focused on injection vulnerabilities. Khoury et al.~\cite{khoury2023secure} conducted an experiment where they generated 21 programs in 5 different programming languages using GPT-3.5. The programs were intentionally diverse, each designed to highlight the risks associated with specific vulnerabilities. Among these, 6 programs targeted injection vulnerabilities. Notably, in every instance, GPT-3.5 failed to do proper input sanitization, resulting in various types of injection vulnerabilities. Similarly, T{\'o}th et al.~\cite{toth2024llms} focused on the security of 2500 PHP websites, generated by GPT-4. Using both dynamic and static scanners, manual code verification, and penetration testing, they identified various vulnerabilities, particularly concerning the security of file upload functions, as well as SQL injections and XSS.

The studies frequently identified the following injection-related CWEs: CWE-79: Improper Neutralization of Input During Web Page Generation (``Cross-site Scripting'')~\cite{liu2024no,pearce2022asleep,siddiq2022securityeval,siddiq2023generate,hajipour2024codelmsec,he2023large,he2024instruction,gong2024well,jamdade2024pilot, aydin2025security,majdinasab2024assessing}, and CWE-89: Improper Neutralization of Special Elements used in an SQL Command (``SQL Injection'') 
\cite{liu2024no,pearce2022asleep,he2023large,siddiq2022securityeval,fu2023security,he2024instruction,jamdade2024pilot, aydin2025security,majdinasab2024assessing}.

\subsection{Memory Management}
This category includes vulnerabilities related to memory management, identified across most of the included studies. These vulnerabilities include overflow vulnerabilities, such as buffer and integer overflow, mostly caused by the lack of input sanitization. Moreover, this category includes null pointer dereference and use-after-free memory allocation vulnerabilities. These issues arise due to improper handling of memory references and pose significant risks to system security. 
Although~\cite{liu2024no, khoury2023secure, perry2023users, sandoval2023lost, pearce2022asleep, hajipour2024codelmsec, he2023large, siddiq2022securityeval, tihanyi2025secure, he2024instruction} all identified memory vulnerabilities in code produced by LLMs, Liu et al.~\cite{liu2024no} placed special importance on memory management vulnerabilities. They performed a systematic empirical assessment of the quality of code generated using ChatGPT, by solving 728 algorithmic problems. 
Their study highlighted significant issues in memory management within LLM-generated code. Specifically, they found that the majority of vulnerabilities were related to missing null pointer tests, with 91.8\% of the total vulnerabilities attributed to the MissingNullTest query. These code snippets failed to check for null after memory allocation, which can lead to critical security vulnerabilities such as CWE-476: Null pointer dereference, also identified in~\cite{sandoval2023lost,pearce2022asleep,tihanyi2025secure}.


\subsection{File Management}
File management vulnerabilities are mostly caused by improper handling of files and directories. These vulnerabilities can lead to unauthorized access, data corruption, or system compromise. The most common file management vulnerabilities include improper path restrictions, unrestricted file uploads, and insecure file permissions. 
Several studies identified file management vulnerabilities in LLM-generated code~\cite{liu2024no,khoury2023secure,perry2023users,pearce2022asleep,hajipour2024codelmsec, he2023large, siddiq2022securityeval, fu2023security,he2024instruction, gong2024well,aydin2025security,majdinasab2024assessing}. However, He et al.~\cite{he2023large} dedicated more focus to these vulnerabilities, finding that LLMs often produce code that does not properly restrict file paths based on user input, potentially allowing unauthorized file access. This vulnerability is marked as CWE-22: Improper Limitation of a Pathname to a Restricted Directory (``Path Traversal'')~\cite{liu2024no,pearce2022asleep,hajipour2024codelmsec,siddiq2022securityeval,fu2023security,he2024instruction,gong2024well,aydin2025security,majdinasab2024assessing}.

\subsection{Deserialization}
Deserialization vulnerabilities happen when untrusted data is used to reconstruct objects within an application, and can lead to arbitrary code execution, data manipulation, or denial-of-service attacks. These vulnerabilities are especially concerning in systems that use serialization for communication or storage, as they can be exploited if proper validation and sanitization are not implemented.

Several studies recognized CWE-502: Deserialization of Untrusted Data~\cite{liu2024no,siddiq2022securityeval,khoury2023secure,rabbi2024ai,fu2023security,gong2024well,majdinasab2024assessing,he2024instruction}, however, none of them specifically focused on this topic, leaving some open questions for further research. For example, how frequently do LLMs generate insecure code when prompted with tasks targeting deserialization vulnerabilities?
 
\subsection{Sensitive Data Exposure}
\label{subsec:sensitive}
Sensitive data exposure occurs when critical information, such as personal data, financial details, or authentication credentials, are not properly protected and become accessible to unauthorized entities. Sensitive data exposure is often caused by ineffective data encryption, insecure storage, or flaws in data transmission protocols.
Several studies recognized the exposure of sensitive data in the code generated by LLMs~\cite{liu2024no,khoury2023secure, pearce2022asleep,siddiq2023generate,fu2023security,he2024instruction,gong2024well, aydin2025security,majdinasab2024assessing}. Jamdade et al.~\cite{jamdade2024pilot} reported sensitive data exposure in code generated by ChatGPT for web-application user-data handling.
Moreover, Pearce~et~al.~\cite{pearce2022asleep} studied how frequently and why GitHub Copilot generates insecure code by testing it with 89 different scenarios related to high-risk cybersecurity vulnerabilities. 
Besides finding that approximately 40\% of these programs were vulnerable, they showed that Copilot consistently fails when it comes to CWE-200: Exposure of Sensitive Information to an Unauthorized Actor. Similarly, Aydin et al.~\cite{aydin2025security} found that CWE-200 was the most prevalent weakness in LLM-generated JavaScript across 6 LLMs (127 occurrences, 21\% of reported vulnerabilities). The same CWE was also mentioned in~\cite{liu2024no,pearce2022asleep,gong2024well,jamdade2024pilot,majdinasab2024assessing}.


\subsection{Authentication and Authorization}
\label{subsec:authorization}
Authentication and authorization vulnerabilities represent a significant category of security issues, highlighted in 13 out of 21 studies.
\textbf{Authentication} vulnerabilities occur when a system fails to properly verify the identity of users or devices. This kind of vulnerability is mostly caused by weak password requirements, insufficiently protected credentials, or the use of hard-coded credentials. 
Hamer et al.~\cite{hamer2024just} compared the security properties of the code generated by ChatGPT, with answers provided in Stack Overflow. They concluded that ChatGPT-generated code had 20\% fewer vulnerabilities than Stack Overflow snippets, although both produced security flaws. One of the most frequent CWES in both ChatGPT-generated code and Stack Overflow snippets was CWE-798: Use of Hard-coded Credentials. CWE-798 is also reported in~\cite{liu2024no,pearce2022asleep,siddiq2023generate, hamer2024just, fu2023security, gong2024well,firouzi2024time}. In contrast, Majdinasab et al.~\cite{majdinasab2024assessing} evaluated Copilot using multiple targeted CWE scenarios. For CWE-798 specifically, they reported zero vulnerable outputs for the corresponding prompts.

\textbf{Authorization} vulnerabilities arise when a system fails to enforce access control policies properly, which allows users or processes to perform actions beyond their intended permissions. This vulnerability is often caused by improper implementation of access control. The associated CWE is CWE-284: Improper Access Control, as discussed by Asare et al.~\cite{asare2023github}, who conducted a comparative empirical analysis of Copilot-generated code, showing that GitHub Copilot is less likely to produce vulnerable code compared to human developers.

\subsection{Cryptography}
Cryptographic vulnerabilities arise when encryption mechanisms are poorly implemented or completely absent. Common causes include the use of broken or insecure cryptographic algorithms, improper validation of certificates, the use of weak or absent hashing functions, and the reuse or hard-coding of cryptographic keys.
Several studies~\cite{khoury2023secure, siddiq2023generate,siddiq2022securityeval,hamer2024just,rabbi2024ai,fu2023security,he2024instruction,gong2024well,aydin2025security} identified cryptographic weaknesses in LLM-generated code without exploring them in detail.
Perry et al.~\cite{perry2023users} studied how users interact with AI code assistants to solve security-related tasks. 
The study found that participants with access to an AI assistant were more likely to produce incorrect and insecure solutions, compared to a control group.
Specifically, in one cryptography-related task, participants with AI assistance were more likely to use trivial ciphers or fail to authenticate the final output. 

A common cryptographic vulnerability identified in multiple studies is CWE-327: Use of a Broken or Risky Cryptographic Algorithm~\cite{siddiq2022securityeval,rabbi2024ai,he2024instruction,tihanyi2025secure,aydin2025security,firouzi2024time}.

\subsection{Resource Management}
Resource Management vulnerabilities refer to weaknesses in the handling, allocation, and release of system resources that are not strictly related to memory or file systems. These vulnerabilities were also identified across multiple studies as improper control of resources, resource leaks, and improper resource shutdowns.
Hamer et al.~\cite{hamer2024just} identified two prevalent resource management vulnerabilities in ChatGPT-generated code: CWE-404: Improper Resource Shutdown or Release, and CWE-772: Missing Release of Resource after Effective Lifetime. Both CWEs were tied as the second most frequent vulnerabilities, each appearing in 37, out of 216 code snippets. Other studies also highlighted related vulnerabilities, including  CWE-605: Multiple Binds to the Same Port~\cite{rabbi2024ai,gong2024well}, and CWE-664: Improper Control of a Resource Through its Lifetime~\cite{asare2023github}.

In general, resource management vulnerabilities in LLM-generated code were identified by~\cite{khoury2023secure, asare2023github, hamer2024just, rabbi2024ai, fu2022linevul,gong2024well,aydin2025security}, however, none of these studies explored them in depth.

\subsection{Coding Standards}
This vulnerability category includes issues arising when software development practices do not adhere to established standards, leading to inconsistent, unreliable, or insecure code. Such deviations can introduce bugs, security flaws, or system crashes. Common vulnerabilities of this type include logical errors, such as division by zero and incorrect calculations, as well as issues with function calls, such as incorrect number of arguments or incorrect argument types.
Tihanyi et al.~\cite{tihanyi2025secure} conducted a comparative analysis of LLMs to examine how likely they are to generate vulnerabilities when writing simple C programs. They identified division by zero as a frequent issue in the code generated by LLMs, along with the challenges in managing arrays and arithmetic operations.

Additionally, CWE-758: Reliance on Undefined, Unspecified, or Implementation-Defined Behavior is another significant concern discussed by Sandoval et al.~\cite{sandoval2023lost}. Besides the mentioned studies, others focused on this type of vulnerability, but without exploring it in detail, as can be seen in~\cite{liu2024exploring, khoury2023secure,asare2023github,pearce2022asleep,hajipour2024codelmsec,siddiq2022securityeval,toth2024llms,rabbi2024ai,fu2023security,he2024instruction,gong2024well,tihanyi2025secure}.

\subsection{Error Handling}
Error handling vulnerabilities arise when a system or application fails to properly manage exceptional conditions. Poor error handling can lead to various issues, including security vulnerabilities, system crashes, or unreliable software behavior. One common issue in error handling is the failure to check or handle return values from functions or system calls properly.
A few studies focused on error handling vulnerabilities~\cite{sandoval2023lost,rabbi2024ai,fu2023security,gong2024well,aydin2025security}, however, none of these studies specifically addressed this category of vulnerabilities, which presents another topic for future research.
Among the various CWEs, CWE-703: Improper Check or Handling of Exceptional Conditions has been specifically highlighted in the literature as a notable issue~\cite{rabbi2024ai}.

\subsection{LLMs in Code Generation: Balancing Risks and Benefits}
\label{subsec:risksbenefits}

As we described throughout this section, LLMs often produce code with security vulnerabilities~\cite{perry2023users,siddiq2023generate,toth2024llms,fu2023security,tihanyi2025secure}. Interestingly, when ChatGPT fixes human-written code, the revised version often contains more security vulnerabilities compared to code generated from scratch by ChatGPT~\cite{rabbi2024ai}.

Despite these challenges, LLMs remain valuable tools in coding tasks. Research indicates that code generated by ChatGPT generally introduces fewer CWE issues compared to code found on Stack Overflow~\cite{hamer2024just}. Moreover, LLMs do not significantly increase the incidence rates of severe security bugs~\cite{sandoval2023lost}, and tools like GitHub Copilot have been shown to produce code with fewer security vulnerabilities than code written by human developers~\cite{asare2023github}.

Although LLMs may generate vulnerable code, the literature suggests that using more effective prompts can guide them to produce safer code~\cite{liu2024no,khoury2023secure}. Through improved re-prompting, the security issues initially present in LLM-generated code can often be addressed, making LLMs valuable tools for creating secure code when used appropriately. This topic will be explored in more detail in Subsection \ref{subsec:self-produced}.
In practice, the evidence in this section points to a more cautious use of LLMs in software development. Developers should assume that suggestions may contain vulnerabilities and review them carefully before integrating the code. 
 
\section{LLMs for Vulnerability Detection}
\label{sec:detection}

While LLMs can introduce vulnerabilities into the code they generate, the same models also offer opportunities to improve software security, particularly through their use in vulnerability detection.
Security code reviews performed manually require significant time and effort, especially in large-scale open-source projects with many contributions. Hence, the automated tools able to identify security vulnerabilities during code review are highly beneficial~\cite{paul2021security}. 
Among these tools, LLMs emerged as a promising solution for detecting security vulnerabilities. 

In this section, we address the vulnerability detection aspect of RQ2 by exploring how LLMs are used to detect vulnerabilities in code. We first describe the benchmarks and datasets used for evaluation, then summarize the programming languages and the LLMs evaluated in the reviewed studies. We further review how LLMs compare with non-LLM approaches and discuss reported challenges and influencing factors. Finally, we examine the proposed methods for improving effectiveness in this task.
Across the works we review in this section, performance is typically reported in terms of F1 score and recall, sometimes complemented by precision and False Positive (FP) rate. However, the primary studies do not report a consistent set of metrics. Some focus only on F1 or accuracy, others use detection and FP rates or the number of vulnerabilities found, and some adopt task-specific scoring schemes. Together with differences in datasets, benchmarks, and task formulations, this makes direct comparisons of reported performance inconclusive.

\subsection{Benchmarks and Datasets used for LLM-Based Vulnerability Detection}
\label{subsec:benchmarks}

Before examining how effective LLMs are at detecting vulnerabilities, we first outline the benchmarks and datasets on which they are evaluated, as these choices strongly influence the performance that studies report.
As summarized in Table~\ref{tab:rq2_detection_models_languages_datasets} in Appendix~\ref{app:appendix}, many studies rely on existing vulnerability datasets and benchmarks such as SARD, NVD, Juliet, Devign, and BigVul. Several papers also curate subsets of these datasets or combine multiple sources into one evaluation set.

In addition to these adapted datasets, other studies introduce new, named benchmarks and datasets specifically designed for evaluating LLM-based vulnerability detection. For instance, Ding et al.~\cite{ding2024vulnerability} introduced the PrimeVul benchmark and demonstrated that performance reported on commonly used benchmarks can be overestimated, leading models to appear more effective than they actually are in more realistic settings.
For example, the 7B StarCoder2 model achieved a 68.26\% F1 score on the popular BigVul dataset but only 3.09\% on PrimeVul. Similarly, Carletti et al.~\cite{carletti2025evaluating} proposed MVFSC, a benchmark of 972 real-world C/C++ functions from production open-source projects, showing that LLMs performance drops substantially when evaluated on real-world codebases. Other works propose additional smaller benchmarks that vary by language/platform and evaluation design~\cite{li2025sv, 10.1145/3590777.3590780, ullah2024llms,kouliaridis2024assessing}.

These datasets and benchmarks cover different languages and difficulty levels, and should be taken into account when interpreting reported LLM performance. Recent benchmark proposals further suggest that commonly used benchmarks may make models seem more effective than they are. This should be kept in mind when drawing conclusions about performance in realistic settings.

\subsection{Programming Languages and Models used for LLM-Based Vulnerability Detection}
\label{subsec:languagesAndModels}

Beyond benchmarks and datasets, LLM-based vulnerability detection is also influenced by the programming languages and models used.

Some studies explicitly examine how LLM performance varies across programming languages. For instance, Lin et al.~\cite{lin2025large} compared vulnerability detection on Java and C/C++ and reported higher performance on Java, with an F1 score of 57.98\% on Java compared to 34.71\% on C/C++. Khare et al.~\cite{khare2025understanding} also reported a similar trend on the Juliet benchmark, where under the same model and prompt setting, the F1 score is 79\% on Juliet Java compared to 70\% on Juliet C/C++.
More broadly, Yu et al.~\cite{yu2025preliminary} showed that vulnerability detection performance varies substantially across programming languages, highlighting limited cross-language generalization.

\begin{figure}[tb]
    \centering
    \begin{minipage}{0.49\linewidth}
        \centering
        \includegraphics[width=\linewidth, trim=4mm 3mm 6mm 4mm, clip]{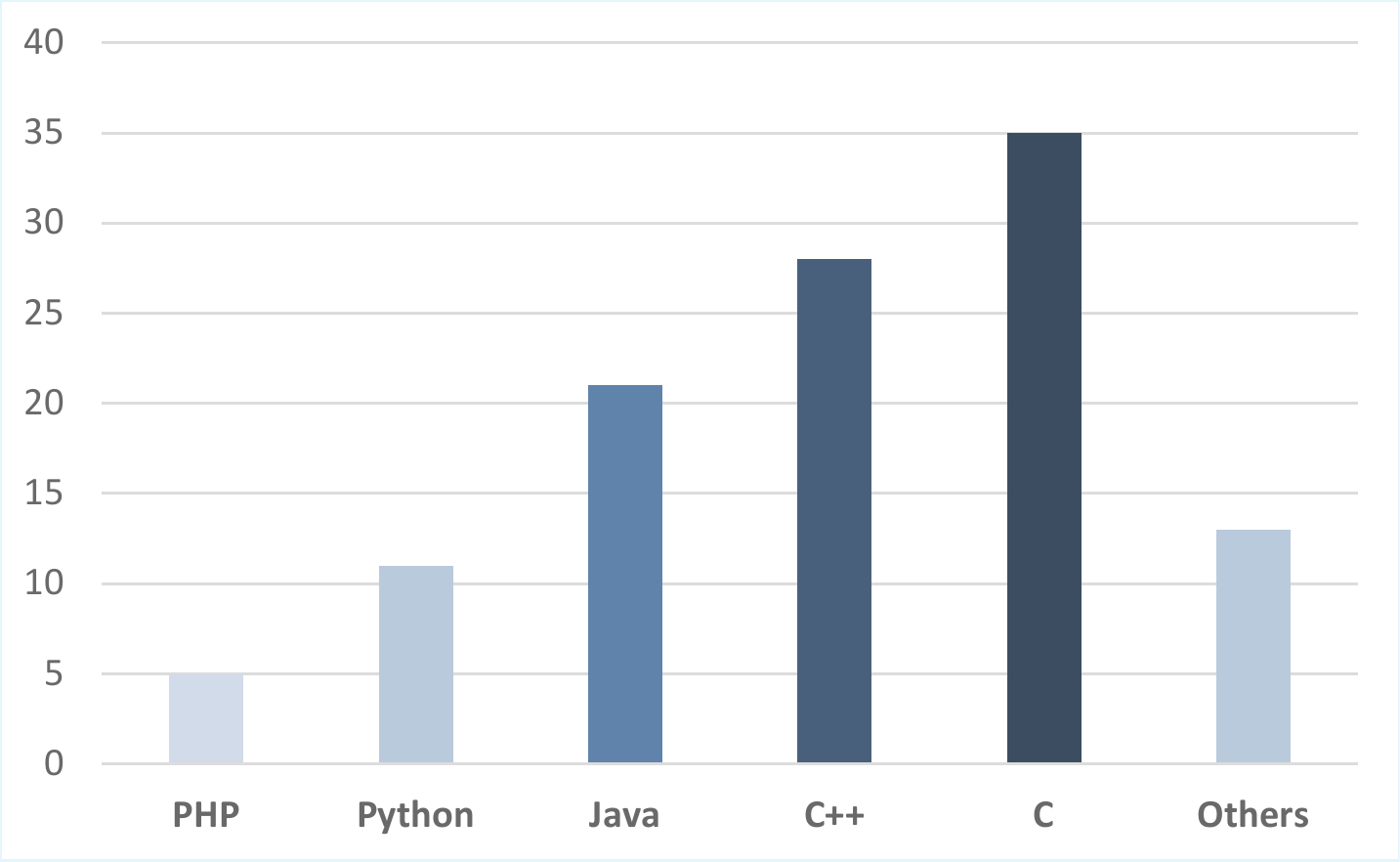}
        \caption{Distribution of programming languages in LLM-based vulnerability detection studies.}
        \label{fig:lang-detection}
    \end{minipage}%
    \hfill
    \begin{minipage}{0.49\linewidth}
        \centering
        \includegraphics[width=\linewidth, trim=4mm 3mm 6mm 4mm, clip]{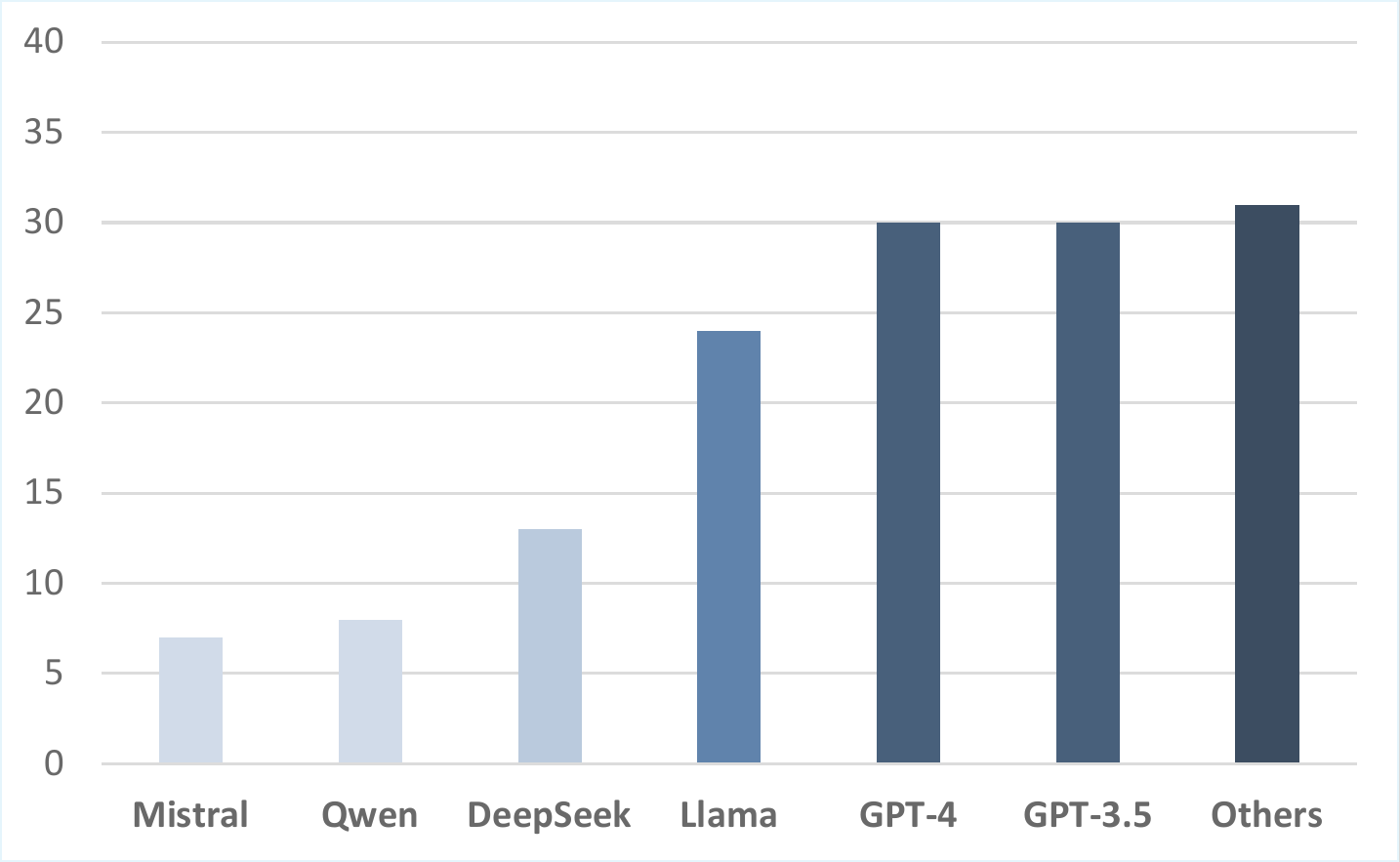}
        \caption{Distribution of LLM families used in vulnerability detection studies.}
        \label{fig:model-detection}
    \end{minipage}
\end{figure}

Most other studies include one or more programming languages in their datasets but do not report language-specific results.
Across the 57 included studies, Figure~\ref{fig:lang-detection} shows that C (35 studies) and C++ (28 studies) are the most frequently evaluated languages. Java, Python, and PHP appear less frequently (21, 11, and 5 studies, respectively), while the remaining languages are grouped as ''Others”. Because most studies evaluate multiple languages, these counts are not mutually exclusive.

Several studies evaluate multiple LLMs side by side~\cite{ding2024vulnerability,guo2024outside,11028575,li2025sv,curto2024can,xia2025beyond,ingemann2024software,gnieciak2025large,mao2025towards,tian2025enhanced,qiu2026rlv}, whereas many other works treat the model choice as just one experimental factor among others.
Figure~\ref{fig:model-detection} summarizes the LLM families evaluated in the reviewed studies. GPT-based models dominate the literature, with both GPT-3.5 and GPT-4 appearing in 30 of the 57 studies. Llama is also widely evaluated (24 studies), while DeepSeek, Qwen, and Mistral appear less frequently (13, 8, and 7 studies, respectively). The remaining low-frequency models are grouped as ''Others” (31 studies). Because most studies evaluate multiple models, these counts are not mutually exclusive.

In conclusion, language and model choices influence LLM-based vulnerability detection performance. Cross-language comparisons report higher performance on Java than on C/C++, but further investigations are warranted for different languages. Overall, studies focus on C/C++ and mainly use GPT-3.5/GPT-4, with a smaller share considering other programming languages and model families. For further details, we refer the reader to Table~\ref{tab:rq2_detection_models_languages_datasets} in Appendix~\ref{app:appendix}

\subsection{Comparison of LLMs with non-LLM Vulnerability Detection Approaches}
\label{subsec:llms_vs_trad}

LLMs show promise in vulnerability detection, but it is essential to compare their performance against other approaches, such as SATs and other ML models (including PLMs).
Table~\ref{tab:traditional-approaches} summarizes non-LLM approaches used as baselines when evaluating LLMs. The upper block lists SATs, while the lower block reports ML-based baselines, including PLMs. 
The Outcome column reports which approach achieved the best performance as defined in each study evaluation.

\begin{table*}[tb]
    \renewcommand{\arraystretch}{1.042}
    \footnotesize
    \centering
    \caption{Non-LLM baselines used to evaluate LLM-based vulnerability detection. Outcome indicates the best-performing approach reported per study, where LLM+SAT denotes a pipeline combining an LLM with a SAT, and LLM/PLM (or LLM/ML) denotes similar performance.}

    \label{tab:traditional-approaches}

    \begin{tabular}{p{0.21\linewidth} p{0.6\linewidth} p{0.08\linewidth}}
        \toprule
        \rowcolor{white}
        \textbf{Reference} & \textbf{Baselines} & \textbf{Outcome} \\
        \midrule

         \multicolumn{3}{l}{\textbf{Static analysis tools (SATs) }} \\[0.2ex]
         \midrule
         
        \rowcolor{lightergray}
        Ozturk et al.~\cite{10.1145/3590777.3590780}* 
            & Betterscan CE, OWASP WAP, PHPcs, Pixy, Progpilot, Psalm, RATS, RIPS, SonarQube, VCG, Horusec 
            & LLM \\

        Tamberg et al.~\cite{tamberg2024harnessing}*
            & CodeQL, SpotBugs 
            & LLM \\

        \rowcolor{lightergray}
        Purba et al.~\cite{purba2023software}*
            & Checkmarx, Flawfinder, RATS 
            & LLM \\

        Gnieciak et al.~\cite{gnieciak2025large}*
            &  SonarQube, CodeQL, SnykCode
            & LLM \\

        \rowcolor{lightergray}
        Wang et al.~\cite{wang2026fine} 
            &  Fortify, SpotBugs
            & LLM \\

        Kouliaridis et al.~\cite{kouliaridis2024assessing}
            &  Bearer, MobSFscan
            & LLM \\

        \rowcolor{lightergray}
        Xia et al.~\cite{xia2025beyond}
            &  CryptoGuard, CogniCryptSAST, SpotBugs
            & LLM \\
            
        Chen et al.~\cite{chen2025vulkiller}
            & Fortify, SpotBugs
            & LLM \\

        \rowcolor{lightergray}
        Li et al.~\cite{li2024llm}
            & CodeQL
            & LLM+SAT \\

        Li et al.~\cite{li2024enhancing}
            & UBITect
            & LLM+SAT \\
            
        \rowcolor{lightergray}
        Mohajer et al.~\cite{mohajer2024effectiveness} 
            & Infer 
            & LLM+SAT \\

        Bakhshandeh et al.~\cite{bakhshandeh2023using} 
            & Bandit, Semgrep, SonarQube 
            & LLM+SAT \\

        \rowcolor{lightergray}
        Xu et al.~\cite{xu2025vulpelican} 
            & Joern 
            & LLM+SAT \\

         Munson et al.~\cite{11028575} 
            &  Semgrep
            & LLM+SAT \\
            
        \midrule

        \multicolumn{3}{l}{\textbf{Machine Learning (ML)-based models}} \\[0.2ex]

        \midrule
        \rowcolor{lightergray}
        Chen et al.~\cite{chen2026gptvd}†
            & Flawfinder, VulDeePecker, SySeVR, SlicedLocator
            & LLM \\

        Khare et al.~\cite{khare2025understanding}† 
            & CodeQL, DeepDFA, LineVul
            & LLM \\

        \rowcolor{lightergray}
        Wen et al.~\cite{wen2024scale}
            & Devign, ReVeal 
            & LLM \\

        Zhang et al.~\cite{zhang2024prompt} 
            & Bugram, CFGNN 
            & LLM \\

        \rowcolor{lightergray}
        Jiao et al. \cite{jiao2025deepvulhunter}
            &  CFGNN, VulACLLM
            & LLM \\

        Zhu et al. \cite{zhu2024detecting}
            &  BiLSTM, AST+GAT
            & LLM \\

        \rowcolor{lightergray}
        Li et al.~\cite{li2025steering}  
        & VulDeePecker, SySeVR, Devign, ReVeal, AMPLE, CasualVul 
            & LLM \\

        Liu et al.~\cite{liu2023software}
            & SySeVR, VulDeePecker
            & LLM \\

        \rowcolor{lightergray}
        Liu et al.~\cite{liu2024exploration}
            & SySeVR, VulDeePecker
            & LLM \\

        Li et al.~\cite{li2025vulnteam}  
        & VulDeePecker, SySeVR, Devign, ReVeal, CasualVul
            & LLM \\

        \rowcolor{lightergray}
        Tian et al. \cite{tian2025enhanced}
            & LineVul, UnixCoder, LineVD
            & LLM \\

        Mao et al.~\cite{mao2025towards} &  
        LineVul, CausalVul, CodeBERT, CodeT5
        & LLM \\

        \rowcolor{lightergray}
         Qiu et al. \cite{qiu2026rlv}
            &  Devign, ReVeal, MGVD, UniXcoder, CodeT5, LineVul
            & LLM \\
        
        Rafiq et al.~\cite{rafique2025beyond} 
        & LR, Multinomial Na\"{\i}ve Bayes, Linear SVC
        & LLM \\

        \rowcolor{lightergray}
        Maligazhdarova et al.~\cite{maligazhdarova2025comparative} & LR, KNN, Bernoulli Na\"{\i}ve Bayes, Decision Tree, RF, MLP, CNN
        & LLM/ML \\

        Zhang et al.~\cite{zhang2024empirical} 
            & VulTeller, Transformer, CodeBERT, GraphCodeBERT, PLBART, CodeT5 
            & ML \\
            
        \rowcolor{lightergray}
        Ni et al.~\cite{ni2024learning} 
            & Devign, ReVeal, IVDetect, LineVul, SVulD 
            & ML \\

        Thapa et al.~\cite{thapa2022transformer} 
            & BiLSTM, BiGRU, BERT, MegatronBERT, DistilBERT, CodeBERT
            & LLM/PLM \\

        \rowcolor{lightergray}
         Zhou et al. \cite{zhou2024large}
            & CodeBERT
            & LLM \\
            
        Shestov et al. \cite{shestov2024finetuning}
            & ContraBERT
            & LLM \\

        \rowcolor{lightergray}
        Guo et al. \cite{guo2024outside}
            & CodeBERT, VulBERTa-MLP variants
            & LLM \\

        Yu et al. \cite{yu2025preliminary}
            & UniXcoder, CodeBERT, CodeT5, CodeT5P, LineVul
            &  PLM\\

        \rowcolor{lightergray}
        Zhang et al.~\cite{zhang2026vultrlm} 
        & UniXcoder, CodeBERT
            &  PLM\\

        Curto et al.~\cite{curto2024can} 
        &  CodeBERT, PolyCoder, NatGen
            & LLM/PLM\\
            
        \bottomrule

 \end{tabular}

\vspace{0.3em}
\makebox[\textwidth][r]{%
  \parbox{0.98\textwidth}{\small
    \textbf{Note:} Studies marked with * showed that LLMs detected more vulnerabilities than SATs, but produced \\a higher FP rate. Studies marked with † use both SATs and ML-based approaches as baselines.
    \\
  }%
}

\end{table*}

In multiple studies, LLM approaches achieve higher vulnerability detection performance than SAT baselines. Wang et al.~\cite{wang2026fine} proposed a pipeline that uses a fine-tuned LLM to detect vulnerabilities and validates suspicious cases with execution-based checks. They reported that their approach outperforms SATs, achieving 95.07\% F1 score, compared to 78.36\% for the best-performing SAT baseline. Kouliaridis et al.~\cite{kouliaridis2024assessing} evaluated 9 LLMs against two SATs. Across 100 vulnerable samples, the best-performing SAT flagged 29 security issues, whereas the best LLM (Code Llama) detected 81.
Xia et al.~\cite{xia2025beyond} reported that task-aware prompting and multi-query self-validation enable GPT-4 to outperform cryptography-focused SATs, improving accuracy over the best SAT baseline by 26.7\%. Chen et al.~\cite{chen2025vulkiller} reported that their proposed pipeline, VulKILLER, achieved better results than SAT baselines, including substantially lower FP rates.

Some studies reported a trade-off such that higher vulnerability detection comes at the cost of higher FP rates. 
Purba et al.~\cite{purba2023software} reported higher detection by LLMs, but with FP rates exceeding 60\%, compared to rates under 45\% for SATs. Ozturk et al.~\cite{10.1145/3590777.3590780} reported that ChatGPT identified 62-68\% of vulnerabilities, while the best-performing SAT detected 32\%, with FP rates of 91\% for ChatGPT and up to 82\% for SATs. Gnieciak et al.~\cite{gnieciak2025large} reported that LLMs performed better overall but exhibited approximately 34–65\% FP, while SATs were generally under 25\%. Tamberg et al.~\cite{tamberg2024harnessing} reached similar conclusions, adding that adaptability and contextual understanding still make LLMs effective for vulnerability detection. We discuss FP in more detail in Subsection~\ref{subsec:issues}.
  

Several studies explored pipelines where LLMs and SATs complement each other to improve vulnerability detection.
Li et al.~\cite{li2024llm} proposed IRIS, a framework that uses an LLM to generate CodeQL taint specifications, runs CodeQL to find vulnerability reports (taint flows), and then uses an LLM to review those reports to filter FP. Compared to using CodeQL alone, IRIS increased detected vulnerabilities by 35\% and reduced FP by 80\%. 
Li et al.~\cite{li2024enhancing} introduced LLift, which runs UBITect with symbolic execution and applies an LLM to undecided cases, improving accuracy and uncovering 4 UBI bugs in the Linux kernel that UBITect alone could not confirm. Xu et al. proposed VulPelican, where LLM queries the SAT for cross-file data flows and function dependencies. The retrieved information is fed back to the LLM, which refines its analysis over multiple rounds.

Some studies use LLMs on top of SAT outputs to confirm reported bugs and discover additional ones.
Mohajer et al.~\cite{mohajer2024effectiveness} applied Infer to detect potential vulnerabilities and then used ChatGPT to re-evaluate the reported warnings. ChatGPT achieved high precision in filtering FP.
Bakhshandeh et al.~\cite{bakhshandeh2023using} found that ChatGPT alone did not outperform SATs, but using it to re-check SAT findings improved F1.
Munson et al.~\cite{11028575} prompted an LLM with code and Semgrep reports to determine whether each reported issue is a true vulnerability or an FP. Semgrep achieved 74.15\% F1, while Semgrep+o1-mini improved this to 86.92\%.

Beyond SATs, researchers also compare LLMs against ML-based baselines, mainly specialized learning-based vulnerability detectors, to assess the effectiveness of LLMs in detecting vulnerabilities. Jiao et al.~\cite{jiao2025deepvulhunter} proposed an LLM-based vulnerability detection framework, and reported that it outperformed ML baselines, achieving up to 75.3\% accuracy compared to 52.0\% for the best-performing ML baseline.
Chen et al.~\cite{chen2026gptvd} reported that their LLM-based framework outperforms all SAT and ML baselines, achieving an F1 score of 92.73\%, compared to 89.53\% for the best-performing baseline. Zhu et al.~\cite{zhu2024detecting} fine-tuned Qwen2-7B and achieved 91.44\% average accuracy, surpassing the BiLSTM baseline (83.39\%). Khare et al.~\cite{khare2025understanding} evaluated 16 LLMs for vulnerability detection and reported that their best LLM-based configuration outperforms the non-LLM baselines in the main F1-based comparisons, reaching 65\% F1.
Additional studies report that LLM-based vulnerability detection outperforms ML-based baselines in their evaluations~\cite{zhang2024prompt,wen2024scale,li2025steering,liu2024exploration,liu2023software,li2025vulnteam,tian2025enhanced,qiu2026rlv}. In contrast, some studies report that specialized ML-based vulnerability detectors achieve higher overall performance than LLMs on the benchmarks they evaluate~\cite{zhang2024empirical,ni2024learning}.

Beyond specialized ML detectors, Rafiqe et al.~\cite{rafique2025beyond} compared fine-tuned LLMs with classical ML classifiers (Logistic Regression, Multinomial Na\"{\i}ve Bayes, and Linear SVC), finding that fine-tuned LLMs achieved higher test accuracy (up to $\approx$84\%) than the strongest ML baseline ($\approx$78\%).
Maligazhdarova et al.~\cite{maligazhdarova2025comparative} reported task-specific results for injection vulnerability detection, when comparing a fine-tuned GPT-4o-mini against multiple classical ML classifiers. For SQL injection, Random Forest achieved 99\% accuracy, whereas for NoSQL injection, a fine-tuned GPT-4o-mini reached 97\% accuracy and outperformed all ML baselines.

Thapa et al.~\cite{thapa2022transformer} evaluated several Transformer-based language models, including both LLMs and PLMs, and found that both model families achieve comparable performance overall and outperform non-Transformer ML baselines. Several other studies use PLMs, mostly from the BERT family, as baselines when evaluating LLMs for vulnerability detection~\cite{zhou2024large,shestov2024finetuning,guo2024outside,mao2025towards,tian2025enhanced,qiu2026rlv}. Overall, these works show that properly tuned or carefully prompted LLMs often outperform these PLM baselines.

However, some studies report that PLM-based detectors outperform LLMs. Yu et al.~\cite{yu2025preliminary} reported an F1 score of 70.75\% for CodeT5P, while the best-performing LLM, QwQ-plus, reached 51.11\%. Zhang et al.~\cite{zhang2026vultrlm} similarly reported that, in their evaluation, the best-performing LLM achieved 58.35\% F1, while MAGNET achieved 66.12\% F1. The authors suggested that this is likely because LLMs are mainly pre-trained for code generation, not for learning vulnerability patterns. Curto et al.~\cite{curto2024can} reported mixed results. Fine-tuned Llama~3 and Code~Llama performed better for detecting vulnerable functions, but CodeBERT was better at pinpointing the vulnerable lines.

In summary, many studies report that LLMs match or outperform SATs for vulnerability detection, but often at the cost of higher FP. However, combining LLMs with SATs frequently improves detection. Comparisons against ML-based baselines and PLM baselines show mixed results, and the best-performing approach varies across studies and evaluation settings.

\subsection{Challenges in Using LLMs for Vulnerability Detection}
\label{subsec:issues}

While LLMs show considerable promise for vulnerability detection, several challenges remain. In this section, we aim to unwrap the challenges reported in the literature. 
Figure~\ref{fig:issues-dot-matrix} presents a mapping between the reviewed studies (rows) and the reported challenges (columns). A cross ($\times$) indicates that the corresponding study reported the given challenge. 
Most studies report FP rates or performance with complex scenarios, while the other two challenges appear less frequently.

\begin{figure}[!tb]
\centering
\begin{tikzpicture}
\begin{axis}[
    width=0.9\linewidth,
    height=9.5cm,                 
    xmin=0.5, xmax=4.5,
    ymin=0.5, ymax=21.5,           
    xtick={1,2,3,4},
    xticklabels={
        False Positives,
        Missing Context,
        Complex Scenarios,
        Token Limits
    },
    x tick label style={align=center},
    ytick={1,...,21},
    yticklabels={
        Ullah~et~al.~\cite{ullah2024llms},
        Purba~et~al.~\cite{purba2023software},
        Ding~et~al.~\cite{ding2024vulnerability},
        \c{C}etin~et~al.~\cite{ccetinempirical},
        Zhang~et~al.~\cite{zhang2024empirical},
        Li~et~al.~\cite{li2023assisting},
        Tamberg~et~al.~\cite{tamberg2024harnessing},
        Wang~et~al.~\cite{wang2026fine},
        Jiao~et~al.~\cite{jiao2025deepvulhunter},
        Kouliaridis~et~al.~\cite{kouliaridis2024assessing},
        Carletti~et~al.~\cite{carletti2025evaluating},
        Zhang~et~al.~\cite{zhang2026vultrlm},
        Xu~et~al.~\cite{xu2025vulpelican},
        Chen~et~al.~\cite{chen2026gptvd},
        Li~et~al.~\cite{li2025sv},
        Xia~et~al.~\cite{xia2025beyond},
        Gnieciak~et~al.~\cite{gnieciak2025large},
        Li~et~al.~\cite{li2025steering},
        Chen~et~al.~\cite{chen2025vulkiller},
        Mao~et~al.~\cite{mao2025towards},
        Zhou~et~al.~\cite{zhou2025ssrfseek}
    },
    y dir=reverse,
    grid=major,
    grid style={dashed,gray!25},
    tick align=inside,
    axis lines*=left,
    enlargelimits=false,
    label style={font=\small},
    tick label style={font=\footnotesize},
]

\addplot[
    only marks,
    mark=x,
    mark size=2.7pt,
    line width=0.7pt,
]
table[row sep=\\]{
x y \\
3 1 \\
1 2 \\
3 3 \\
1 4 \\
4 5 \\
1 6 \\
1 7 \\
1 8 \\ 4 8 \\
1 9 \\
2 10 \\
1 11 \\ 3 11 \\
1 12 \\ 4 12 \\
2 13 \\ 3 13 \\ 4 13 \\
2 14 \\
1 15 \\ 3 15 \\
1 16 \\ 2 16 \\ 3 16 \\ 4 16 \\
1 17 \\
3 18 \\
1 19 \\
3 20 \\
1 21 \\
};

\node[font=\footnotesize,anchor=north] at (axis cs:1,21.45) {13};
\node[font=\footnotesize,anchor=north] at (axis cs:2,21.45) {4};
\node[font=\footnotesize,anchor=north] at (axis cs:3,21.45) {8};
\node[font=\footnotesize,anchor=north] at (axis cs:4,21.45) {5};
\node[font=\footnotesize,anchor=north east] at (rel axis cs:1,0) {\textit{Total}};

\end{axis}
\end{tikzpicture}
\caption{Matrix visualization mapping reported challenges (columns) to reviewed studies (rows) for LLM-based vulnerability detection. A cross ($\times$) denotes that a study reports the corresponding challenge.}
\label{fig:issues-dot-matrix}
\end{figure}

\textbf{FP} represent a key challenge for LLMs, where the model incorrectly flags non-existent vulnerabilities. Some studies report that, while LLMs outperform SATs in vulnerability detection, they also produce higher FP rates~\cite{10.1145/3590777.3590780,tamberg2024harnessing,purba2023software,gnieciak2025large}. Çetin et al.~\cite{ccetinempirical} found that GPT-4 exhibited an FP rate of 63\%, with other models reaching up to 97\%. High FP rates are also reported in other works~\cite{carletti2025evaluating,li2025sv}. Several studies propose techniques to mitigate FP when using LLMs as vulnerability detectors~\cite{li2023assisting,wang2026fine,jiao2025deepvulhunter,zhang2026vultrlm,xia2025beyond,chen2025vulkiller,zhou2025ssrfseek}.

Kouliaridis et al.~\cite{kouliaridis2024assessing} observed that LLM-based vulnerability detection is often limited by \textbf{missing context}. Missing context refers to cases where the prompt lacks essential information needed to judge whether code is vulnerable. Xu et al.~\cite{xu2025vulpelican} showed that LLM-only analysis can miss vulnerabilities when relevant information spans multiple files (e.g., call chains, data flows, dependencies) and is omitted from the prompt. Chen et al.~\cite{chen2026gptvd} reported similar failures when key code details (e.g., variable values or size information) are missing. Xia et al.~\cite{xia2025beyond} further found that limited class-level context can cause LLMs to miss cryptographic misuses that require multi-hop reasoning.

Beyond missing context, LLMs also struggle in \textbf{complex scenarios}, where the relevant information may be available but is harder to use effectively. Li et al.~\cite{li2025sv} found degraded performance in scenarios that require flow-aware (control/data-flow) reasoning. Xu et al.~\cite{xu2025vulpelican} argue that vulnerabilities often involve cross-function call chains and dependencies, making single-function LLM-only detection unreliable. Xia et al.~\cite{xia2025beyond} observed similar performance drops on cryptographic misuses requiring inter-procedural, multi-hop call-chain reasoning.

These complexity-related challenges become more pronounced in real-world settings. Carletti et al.~\cite{carletti2025evaluating} showed that even after fine-tuning, performance drops substantially on real-world codebases not seen during training. Li et al.~\cite{li2025steering} further observed a localization challenge in real-world programs, where vulnerable segments may constitute only a small portion of a function, making them harder for LLMs to detect. Overall, several studies agree that current LLM-based approaches still fall short for vulnerability detection in complex, real-world scenarios~\cite{ullah2024llms,ding2024vulnerability,mao2025towards}.

Token limitations constrain how much code and context can be provided to an LLM during vulnerability detection. Xu et al.~\cite{xu2025vulpelican} identified this challenge when prompting LLMs with SAT outputs. Similarly, Xia et al.~\cite{xia2025beyond} found that performance degrades on longer inputs, sometimes yielding low-quality or even empty outputs near the context limit. Related work reports similar issues and proposes mitigation strategies~\cite{zhang2024empirical, wang2026fine,zhang2026vultrlm}.

The literature highlights several recurring challenges when using LLMs for vulnerability detection. Some of the reviewed works propose techniques to address these challenges~\cite{li2023assisting,zhang2024empirical, wang2026fine,jiao2025deepvulhunter,zhang2026vultrlm,chen2025vulkiller,zhou2025ssrfseek}, which we discuss in more detail in Subsection~\ref{subsec:improving}.
Despite the challenges we highlighted, most of the studies agree that LLMs can be effective tools for vulnerability detection.

\subsection{Factors Influencing LLMs Performance}

\label{subsec:factors}

Despite the remarkable capabilities of LLMs in vulnerability detection tasks, their effectiveness is influenced by various factors, including the prompts used (further explored in Section \ref{sec:prompting}) and the state of the code being analyzed.

Yu et al.~\cite{yu2025preliminary} reported that detection performance varies across programming languages, indicating that the language of the analyzed code is an important factor for vulnerability detection. Jiao et al.~\cite{jiao2025deepvulhunter} found that detection outcomes are impacted by code length. In their analysis, Llama models produced more FP on longer code, while DeepSeek models showed the opposite tendency. Consistent with this, Lin et al.~\cite{lin2024evaluating} observed that some models degrade as input length increases, and Zhou et al.~\cite{zhou2024large} similarly reported degraded accuracy and increased FP for inputs longer than 3000 characters. Beyond length, Sovrano et al.~\cite{sovrano2025large} showed that performance is sensitive to the position of the vulnerable code within the input. Even when an entire file fits within the model context window, LLMs are more likely to miss vulnerabilities located toward the end of large files.

Taken together, these studies show that LLM-based vulnerability detection is highly sensitive to factors beyond the model itself, including programming language, code length, and the position of the vulnerability within the input. These factors should be considered when using LLMs for vulnerability detection in practice.

\subsection{Improving LLM-Based Vulnerability Detection}
\label{subsec:improving}

In light of the limitations and influencing factors we discussed in the previous subsections, many studies explored different ways to enhance LLMs ability to detect vulnerabilities.

One direction for improving LLM-based vulnerability detection is to \textbf{pre-process the code} before feeding it to the model to enrich the input. Thapa et al.~\cite{thapa2022transformer} translated code into vectorized inputs for vulnerability analysis, where GPT-2 Large achieved 95.51\% F1. Wen et al.~\cite{wen2024scale} proposed SCALE, which builds an AST and augments nodes with short auto-generated comments, reaching around 65\% F1. Anbiya et al.~\cite{anbiya2025java} evaluated AST, CPG, and combined AST+CPG representations and concluded that CPG-only inputs were most effective overall.

Other works address long inputs and token limits by splitting code into smaller units while trying to preserve semantics. Zhang et al.~\cite{zhang2024empirical} used overlapping sliding windows with a right-forward embedding strategy to enrich context. Wang et al.~\cite{wang2026fine} used CFG-based semantic segmentation and CFG sub-paths to mitigate token limitations, and applied targeted execution checks to reduce FP. Zhang et al.~\cite{zhang2026vultrlm} decomposed long functions into AST sub-trees and enriched them with LLM-generated comments. Sovrano et al.~\cite{sovrano2025large} improved file-level detection by chunking large files and reported over 37\% average recall improvement, while noting that reduced global context may increase FP. Saimbhi et al.~\cite{saimbhi2024vulnerai} applied function-level segmentation to PHP snippets and prompted an LLM per function for detection and CWE labeling, reporting zero FP but only 68\% recall.

Beyond input-level pre-processing, some studies focus on reducing FP by using LLMs in a multi-step filtering and validation process that checks vulnerability candidates produced by SATs or an initial LLM pass.
Li et al.~\cite{li2023assisting} used ChatGPT to summarize and refine static-analysis warnings, helping developers prioritize issues and eliminating most FP in their experiments. Jiao et al.~\cite{jiao2025deepvulhunter} similarly targeted FP reduction with their multi-round LLM-based vulnerability detection framework, DeepVulHunter. Chen et al.~\cite{chen2025vulkiller} combined Code Property Graph (CPG)-based structural and data/control-flow analysis with LLM-generated Proof-of-Concept exploit inputs (PoCs) and runtime validation to confirm vulnerabilities, reducing the FP rate to 4.3\%. Zhou et al.~\cite{zhou2025ssrfseek} combined PHPJoern-based CPG taint analysis with an LLM to validate Server-Side Request Forgery (SSRF) candidates by checking user control over URL components (e.g., host/authority and path) and path feasibility, reducing 12 taint paths to 7 confirmed vulnerabilities.

Other studies focus on improving the models to enhance vulnerability detection. Li et al.~\cite{li2025steering} proposed an inference-time activation steering approach. They computed a ''vulnerability steering vector” from representation differences between paired vulnerable and patched functions, and injected it into selected transformer layers during inference to bias the model toward vulnerability-related semantics. With this approach, their steered LLMs outperformed ML-based vulnerability detectors, achieving an F1 score of up to 96.25\%.

\begin{figure}[tb]
    \centering
    \includegraphics[width=0.55\columnwidth]{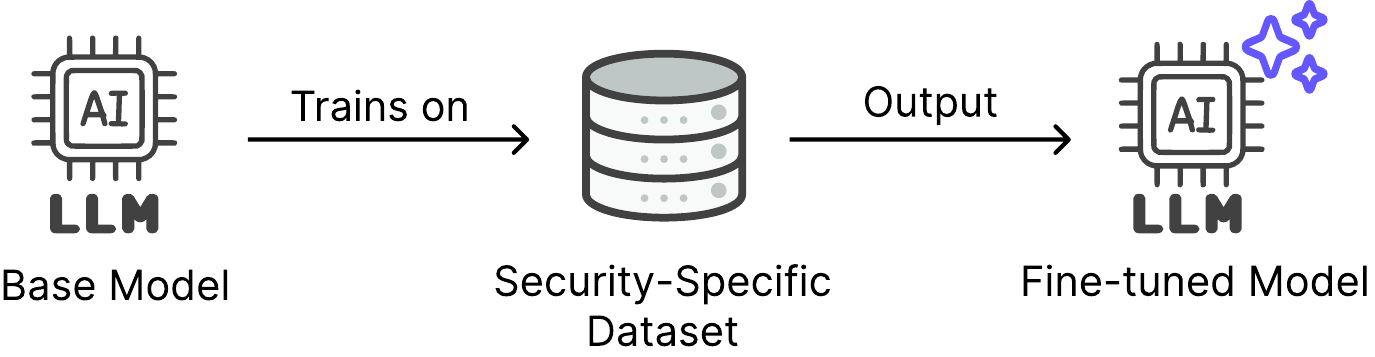}  
    \captionsetup{justification=centering}
    \caption{Fine-tuning an LLM on the security-specific dataset.}
    \label{fig:finetuning}
 \end{figure}

Several researchers explored \textbf{fine-tuning} to improve vulnerability detection. As illustrated in Figure~\ref{fig:finetuning}, fine-tuning adapts a pre-trained model to a security task by further training on a security-specific dataset. Guo et al.~\cite{guo2024outside} reported that a fine-tuned CodeLlama-7B achieved 97\% F1 on their dataset, while the best-performing model without fine-tuning achieved about 51\% F1.  Mechri et al.~\cite{mechri2025secureqwen} fine-tuned Qwen and enabled long-context analysis of larger codebases, achieving F1 scores ranging from 84\% to 99\%.
Curto et al.~\cite{curto2024can} fine-tuned Llama~3 and Code~Llama for vulnerability detection, achieving up to 95.3\% F1 for function-level detection and up to 69.63\% Top-10 accuracy for line-level localization on BigVul.
Ibanez-Lissen et al.~\cite{IBANEZLISSEN2025104125} fine-tuned Gemma-2B for multi-class CWE detection, achieving 95.9\% F1 for binary vulnerability detection.
Zhu et al.~\cite{zhu2024detecting} fine-tuned Qwen2-7B for CWE-level Java vulnerability detection and reported strong performance on their curated dataset.
Shestov et al.~\cite{shestov2024finetuning} fine-tuned WizardCoder for vulnerability detection and reported an F1 score of 71\%.
Wang et al.~\cite{wang2026fine} similarly reported gains from fine-tuning over the base model.

Beyond standard fine-tuning, Zhang et al.~\cite{zhang2024empirical} compared two fine-tuning setups for vulnerability localization, a classification approach that labels each line or statement as vulnerable or not, and a generative approach that predicts vulnerable line numbers and statements. The authors found the classification setup stronger, reporting an F1 score of 82.0\% with CodeLlama-7B, compared to 44.8\% for the generative setup. Moreover, Mao et al.~\cite{mao2025towards} extended fine-tuning by proposing a framework named LLMVulExp, which uses a teacher LLM to generate training explanations for each code sample and fine-tunes a student model to perform vulnerability detection and produce natural-language explanations. In their evaluation, LLMVulExp achieved an F1 score of up to 98.8\%.

Li et al.~\cite{li2025vulnteam} proposed VulnTeam, which extends single-model fine-tuning with role-specialized tuning and aggregation. They fine-tune 5 LLM ``experts” on different vulnerability-feature categories and instruction-tune a lightweight ``leader” model to combine the experts’ analyses into the final decision. VulnTeam outperformed single-model fine-tuning baselines, achieving up to 96.02\% F1.

Despite these improvements, Guo et al.~\cite{guo2024outside} cautioned that gains are largely limited to the training tasks, and that labeling errors in current datasets can compromise training and performance. 
Ding et al.~\cite{ding2024vulnerability} reported that fine-tuning StarCoder2-7B improves F1 from 3.09\% to 18.05\%, but the models still miss many vulnerabilities. Carletti et al.~\cite{carletti2025evaluating} further showed that even after fine-tuning CodeLlama-7B, performance drops substantially on MVFSC, a real-world test set from open-source projects not seen during training.
Some papers also include fine-tuning \cite{maligazhdarova2025comparative, purba2023software, thapa2022transformer, curto2024can,rafique2025beyond,anbiya2025java}. However, in these works, fine-tuning is typically used as part of a broader empirical evaluation or system setup (often for only a subset of the evaluated models), rather than being the primary focus.

In conclusion, pre-processing code, combining LLMs with SATs, and adapting models through fine-tuning all raise vulnerability detection scores on the reported benchmarks. At the same time, prior work notes that these gains are largely limited to the tasks and datasets used for training and evaluation, and that even fine-tuned models still miss many vulnerabilities. 
\section{LLMs for Vulnerability Fixing} 
\label{sec:fixing}

Developers often introduce security vulnerabilities in their code, which, if exploited, can compromise system and application security. 
While prior work has shown that LLMs can be effective at detecting vulnerabilities, their capabilities do not end there. Beyond identifying vulnerabilities, LLMs can assist in providing the right fixes, significantly reducing developers' workloads~\cite{pearce2023examining}. In this section, we address the vulnerability fixing aspect of RQ2 by exploring the use of LLMs for fixing vulnerabilities. We first summarize the datasets, programming languages, and models. Then, we discuss challenges LLMs encounter when fixing vulnerabilities and fine-tuning approaches proposed to address them.

\subsection{Programming Languages, Models, and Datasets for LLM-Based Vulnerability Fixing} 
\label{subsec:languagesModelsFixing}

Before delving into the reviewed studies, it is worth noting that the literature evaluates LLM-based vulnerability fixing under different experimental setups. The models, programming languages, and datasets used in each vulnerability fixing study are listed in detail in Appendix~\ref{app:appendix} (Table~\ref{tab:rq2_fixing_models_languages_datasets}). 

Overall, most vulnerability fixing evaluations focus on C and C++, with fewer studies considering Java and other higher-level languages. Although C and C++ dominate existing evaluations, Wang~et~al.~\cite{wang2025evaluation} performed a cross-language comparison and showed that these languages yield the lowest vulnerability fixing success rates, suggesting that generating correct fixes is more difficult for C/C++ than for other languages such as Go, which achieved the strongest performance.

Across the reviewed vulnerability fixing studies, GPT-family models are analyzed most frequently, while families such as Claude, Codex, LLaMA, and Mistral appear less frequently. Some studies also compare models side by side. For instance, Braconaro et al.~\cite{braconaro2024dataset} evaluated GPT-4o, Gemini 1.5 Flash, and Gemini in Android Studio for vulnerability fixing, reporting 93.8\% accuracy for GPT-4o, compared to 83.8\% and 83.5\% for the two Gemini models, respectively. Similarly, Le et al.~\cite{le2024study} compared ChatGPT and Bard, reporting repair accuracies of 71.66\% and 68.33\%, respectively.

Evaluation of vulnerability fixing relies on a mix of real-world CVE-based benchmarks and curated datasets, most commonly ExtractFix, CVEFixes, Big-Vul, and Vul4J, alongside manually constructed evaluation sets. Braconaro et al.~\cite{braconaro2024dataset} proposed a dataset of 272 security best-practice violations paired with ground-truth repairs. Similarly,
de-Fitero-Domínguez et al.~\cite{de2024enhanced} constructed a refined repair dataset by removing train-test overlap to support more reliable evaluation.

Overall, the vulnerability fixing literature remains centered on GPT-family models and C/C++ code, with fewer studies covering other model families and programming languages. LLM-based vulnerability fixing performance is also evaluated on a diverse set of benchmarks and datasets, which further limits direct comparisons across reported results.

\subsection{Challenges in Using LLMs for Vulnerability Fixing}
\label{subsec:challengesFixing}

The literature on LLM-based vulnerability fixing also identifies several challenges that affect the reliability and effectiveness of generated repairs.

A key challenge in LLM-based vulnerability fixing is handling \textbf{context-sensitive, non-local repairs}. Zhang et al.~\cite{zhang2024evaluating} show that LLMs are effective at fixing localized memory-related issues in C/C++ code, but struggle when repairs depend on external functions, data structures, or complex program logic. However, providing additional contextual information can improve fixing performance. Pearce et al.~\cite{pearce2023examining} similarly report that, while LLMs can fix synthetic and hand-crafted vulnerabilities, performance degrades on real-world code in zero-shot settings due to insufficient context and high sensitivity to prompt formulation and natural language variations. Wang et al.~\cite{wang2026spvr} concluded that their proposed approach can still fail when the correct fix depends on variables, constants, or function names not present in the available code context. Liu et al.~\cite{liu2024exploring} reported similar context-related failures in vulnerability fixing, where missing details can lead ChatGPT to infer or guess information (e.g., a required variable), resulting in an incorrect patch.
  
A further challenge is the \textbf{reliability} of generated fixes, including their \textbf{consistency} across repeated runs. Existing studies reported non-deterministic outcomes across repeated runs and hallucinated or unsupported fixes that introduce elements not present in the provided code or context~\cite{antal2025identifying,braconaro2024dataset}. 
Tihanyi et al.~\cite{tihanyi2025new} observed repairs that appear correct but fail to resolve the vulnerability, such as trivial bound adjustments (e.g., adding $+1$ to an upper bound). 

Another major challenge is that generated fixes may be \textbf{incomplete}, even when they appear reasonable. Tao et al.~\cite{tao2025adapting} note that fix suggestions can miss edge cases or overlook security implications, resulting in partial fixes. Similarly, Tihanyi et al.~\cite{tihanyi2025new} report that repair difficulty varies across vulnerability types, indicating that some classes of issues are harder to fix than others. Moreover, Firouzi et al.~\cite{firouzi2024time} concluded that LLM-based vulnerability fixing is currently most reliable when integrated into a human-in-the-loop workflow. Security improvements often require targeted feedback and manual line-by-line validation rather than one-shot repair prompts.

In summary, the literature highlights that LLM-based vulnerability fixing is sensitive to context and prompt formulation. It is further limited by inconsistent outputs and the risk of incomplete repairs, and reliable security fixes still depend on careful human verification.

\subsection{Improving LLM-Based Vulnerability Fixing}
\label{subsec:improvingFixing}

Given the aforementioned challenges encountered by LLMs in generating reliable, consistent, and complete fixes, several studies have explored various strategies to improve their performance.

A common approach is fine-tuning LLMs, a process described earlier in Section~\ref{sec:detection}. For instance, de-Fitero-Dominguez et al.~\cite{de2024enhanced} fine-tuned Code Llama and Mistral for vulnerability fixing and report that the fine-tuned models achieve up to 23\% and 26\% Perfect Prediction, respectively, on a refined evaluation set without training–test overlap.
Moreover, Wu et al.~\cite{wu2023effective} evaluated LLMs for fixing Java vulnerabilities on two proposed benchmarks, comparing off-the-shelf and fine-tuned variants. Codex fixed 20.4\% of the vulnerabilities without fine-tuning, while the best fine-tuned model, InCoder, fixed 18\%, which corresponds to an 8\% relative improvement over its non-fine-tuned version and was largely limited to simple edits rather than complex CWEs.  Finally, Wang et al.~\cite{wang2025evaluation} investigated fine-tuning for multilingual vulnerability fixing and reported that GPT-4o increased from 26.89\% EM to 28.71\% EM after fine-tuning.

Another approach improves LLM-based vulnerability fixing without changing the model. Instead, it relies on pre-processing code into structured inputs. Wang et al.~\cite{wang2026spvr} propose SPVR, which parses code into an AST and uses simple, rule-based signals to generate targeted prompts and a simplified code snippet for the LLM. With this approach, GPT-4 repair success increased from 65/547 to 143/547 vulnerabilities (pass@3), and CodeBLEU improved from 30.25\% to 41.09\%.

A different strategy focuses on combining context reduction with LLM-based validation. Nong et al.~\cite{nong2025appatch} proposed APPATCH, which reduces the input code by extracting only the parts that influence the vulnerable line. It then prompts the LLM to explain the cause of the vulnerability and generate multiple patch candidates. Finally, APPATCH uses other LLMs to review the candidates and keeps only patches that appear to both fix the vulnerability and avoid breaking program behavior. On ExtractFix APPATCH reached 90\% recall and 68.41\% F1 score.

Several fixing pipelines augment LLM-based patch generation with external non-LLM feedback (e.g., compiler/tests, verifier counterexamples) to improve reliability and enable iterative refinement of generated patches \cite{tihanyi2025new,10.1145/3664646.3664770}. In this direction, Kim et al.~\cite{kim2025logs} proposed SAN2PATCH, which uses crash logs from AddressSanitizer and UBSan to guide localization and patch generation, and validates candidate patches through an automated build-and-test loop.

These studies indicate that fine-tuning, code pre-processing and external patch validation can improve vulnerability fixing. However, overall the gains are often small, and the benefits are mostly limited to simple, localized fixes.
 
\section{LLMs for Detecting and Fixing Vulnerabilities in a Unified Process}
\label{sec:detectionAndFix}


In Section~\ref{sec:detection} and Section~\ref{sec:fixing}, we discussed using LLMs as tools for handling vulnerability detection and vulnerability fixing separately. 
In this section, we explore an integrated approach in which LLMs detect and fix security vulnerabilities in a unified process, including identifying vulnerabilities in code they generate themselves and resolving them through re-prompting.
Last, we cover fine-tuning for enhancing LLMs performance in a unified pipeline for vulnerability detection and fixing.
The models, programming languages, and datasets used in the reviewed studies that combine vulnerability detection and fixing are summarized in Appendix~\ref{app:appendix}, Table~\ref{tab:rq2_detection_fixing_models_languages_datasets}.
Some studies discussed in this section are instead listed in Appendix~\ref{app:appendix}, Table~\ref{tab:rq1_models_languages_datasets}, as they also address RQ1.

\subsection{LLMs for Detecting and Fixing Vulnerabilities in Self-produced Code}
\label{subsec:self-produced}

As we discussed in Section~\ref{sec:indroducedVulnerabilities}, code generated by LLMs may contain security vulnerabilities. However, several studies show that it is possible to re-prompt LLMs to identify and fix those vulnerabilities.
  
Liu et al.~\cite{liu2024no} demonstrated promising results in fixing security vulnerabilities using ChatGPT, with over 89\% of vulnerabilities successfully mitigated during a multi-round fixing process. In this approach, Liu et al. interacted with ChatGPT over several rounds, providing feedback on the generated code and asking for improvements.
Similarly, Gong et al.~\cite{gong2024well} showed that LLMs often produce vulnerable code, with 75\% of the tested code classified as insecure. They further found that advanced LLMs can fix up to 60\% of insecure code from other LLMs, but perform poorly on their own code, a phenomenon named \textit{self-repair blind spots}. To address this, the authors developed a tool that uses LLMs in an iterative fix process supported by semantic analysis, improving the security of LLM-generated code. Finally, Khoury et al.~\cite{khoury2023secure} instructed ChatGPT to generate programs targeting certain vulnerabilities, then analyzed the code for evident vulnerabilities. If present, they re-prompted ChatGPT to identify them and produce a safer version. The study showed ChatGPT could fix several vulnerabilities when re-prompted, but still required significant human guidance.

Overall, multi-round re-prompting can mitigate many vulnerabilities, but LLMs still struggle with \textit{self-repair blind spots} and typically require human guidance or additional analysis.

\subsection{Fine-tuning for Vulnerability Detection and Fixing}
\label{subsec:fineTuningDetectionAndFixing}

To improve vulnerability detection and fixing as a unified process, fine-tuning can be applied to adapt LLMs to security-relevant tasks, following the process described earlier in Section~\ref{sec:detection}.

Zhang et al.~\cite{zhang2024effectiveness} fine-tuned LLMs for GitHub Actions workflow tasks and evaluated defect detection and repair, where defects include both syntactic errors and security issues such as code injection vulnerabilities.
Focusing on vulnerability detection and fixing, they found that fine-tuning improved the detection of code injection vulnerabilities, achieving up to 98.82\% F1 with fine-tuned GPT-3.5, but reported that LLMs were not effective at fixing code injection vulnerabilities.

Fine-tuning can make LLMs highly effective at detecting vulnerabilities, but it does not reliably fix them, so outputs should be carefully reviewed before deployment. 
\section{Prompting strategies for detecting and fixing vulnerabilities}
\label{sec:prompting}

While LLMs show promise in detecting and fixing vulnerabilities, their effectiveness is not solely determined by model architecture or fine-tuning, it also depends on prompt design. Well-crafted prompts can guide LLMs towards more accurate solutions, while incomplete or ambiguous prompts may result in ineffective solutions~\cite{zhang2024empirical}.
In this section, we address RQ2.1 by exploring the impact of various prompting techniques on the detection and fixing performance of LLMs. Table~\ref{tab:prompt} summarizes the different prompting techniques and corresponding studies.

\begin{table}[tb]
\centering
\renewcommand\arraystretch{1.4}
\caption{Prompt engineering techniques used in the studies for vulnerability detection and fixing.}
\label{tab:prompt}
\footnotesize
\begin{tabular}{>{}p{0.15\linewidth}p{0.80\linewidth}}

\rowcolor{white}
\textbf{Technique} & \textbf{References} \\
\toprule
    Zero-shot & 
    \cite{pearce2023examining}, \cite{mohajer2024effectiveness}, \cite{ullah2024llms},  
    \cite{tamberg2024harnessing}, \cite{zhang2024empirical},
    \cite{liu2024exploring},
    \cite{zhang2024effectiveness}, \cite{ni2024learning},
    \cite{10556123}, 
    \cite{yu2025preliminary}, 
    \cite{kouliaridis2024assessing},
    \cite{carletti2025evaluating},
    \cite{xu2025vulpelican},
    \cite{chen2026gptvd},
    \cite{11028575},
    \cite{li2025sv},
    \cite{xia2025beyond},
    \cite{ingemann2024software}, 
    \cite{wang2025evaluation},
    \cite{antal2025identifying}, 
    \cite{braconaro2024dataset}, 
    \cite{firouzi2024time}, 
    \cite{fu2023chatgpt}, 
    \cite{nong2025appatch}, 
    \cite{lin2024evaluating}, 
    \cite{liu2023software}, 
    \cite{trad2025manual}, 
    \cite{khare2025understanding}, 
    \cite{huynh2025detecting}, 
   \cite{qiu2026rlv}
    \\

    \cellcolor{lightergray} Few-shot  & \cellcolor{lightergray} 
    
    \cite{mohajer2024effectiveness}, 
    \cite{ullah2024llms}, 
    \cite{tamberg2024harnessing}, 
    \cite{ding2024vulnerability}, \cite{zhang2024empirical}, \cite{zhang2024effectiveness}, 
    \cite{liu2024exploring},
    \cite{carletti2025evaluating},
    \cite{chen2026gptvd},
    \cite{sovrano2025large},
    \cite{wang2025evaluation}, 
    \cite{fu2023chatgpt}, 
    \cite{nong2025appatch}, 
    \cite{liu2023software}, 
    \cite{rafique2025beyond}, 
    \cite{trad2025manual}
    \\

    In-context & 
    \cite{zhang2024empirical},
    \cite{li2024llm}, 
    \cite{ni2024learning},
    \cite{10556123}, 
    \cite{le2024study}, 
    \cite{jiao2025deepvulhunter},
    \cite{chen2026gptvd},
    \cite{li2025sv}, 
    \cite{antal2025identifying}, 
    \cite{wang2026spvr}, 
    \cite{khan2025code}, 
    \cite{nong2025appatch}, 
    \cite{kim2025logs}, 
    \cite{chen2025vulkiller}, 
    \cite{liu2024exploration}, 
    \cite{liu2023software}, 
    \cite{rafique2025beyond}, 
    \cite{khare2025understanding}, 
    \cite{qiu2026rlv}
    \\

    \cellcolor{lightergray} Chain-of-Thought & \cellcolor{lightergray}
    \cite{zhang2024prompt}, \cite{mohajer2024effectiveness}, 
    \cite{pearce2023examining},  \cite{ullah2024llms},  \cite{tamberg2024harnessing},  \cite{zhang2024empirical},   \cite{zhang2024effectiveness}, \cite{ni2024learning}, 
    \cite{10556123}, 
    \cite{jiao2025deepvulhunter},
    \cite{carletti2025evaluating},
    \cite{chen2026gptvd},
    \cite{11028575}, 
    \cite{ingemann2024software}, 
    \cite{tao2025adapting}, 
    \cite{nong2025appatch}, 
    \cite{kim2025logs}, 
    \cite{mao2025towards}, 
    \cite{liu2024exploration}, 
    \cite{rafique2025beyond}, 
    \cite{trad2025manual}, 
    \cite{huynh2025detecting}, 
    \cite{tian2025enhanced}, 
    \cite{zhou2025ssrfseek}, 
    \cite{qiu2026rlv}
    \\

    Task-oriented  & 
    \cite{ullah2024llms}, 
    \cite{zhou2024large}, \cite{bae2024enhancing},
    \cite{wang2026fine}, 
    \cite{xia2025beyond}\\

    \cellcolor{lightergray} Role-oriented & \cellcolor{lightergray} 
    \cite{ullah2024llms}, 
    \cite{zhou2024large}, 
    \cite{ccetinempirical}, 
    \cite{zhang2024empirical},
    \cite{yang2024dlap}, \cite{mohajer2024effectiveness}, \cite{he2023large}, 
    \cite{guo2024outside}, \cite{ni2024learning}, \cite{zhang2024prompt}, \cite{liu2024exploring}, 
    \cite{wen2024scale},
    \cite{wang2026fine},
    \cite{zhang2026vultrlm},
    \cite{xia2025beyond},
    \cite{gnieciak2025large},
    \cite{wang2025evaluation}, 
    \cite{lin2024evaluating}, 
    \cite{rafique2025beyond}, 
    \cite{li2025vulnteam}
    \\
    
\bottomrule

\end{tabular}
\end{table}

\subsection{Zero-shot Prompting}

Zero-shot prompting is a prompting technique in which a model is asked to perform a task using only an instruction or general guideline, without any task-specific examples in the prompt~\cite{NEURIPS2022_8bb0d291}. We provide an example of zero-shot prompting in Figure~\ref{fig:zero-shot}, demonstrating how the user provides instructions without examples. 
Zero-shot prompting has been explored by most of the studies covered in the remainder of the section. Although effective in some scenarios, zero-shot prompting was generally outperformed by other approaches and used only as a baseline for comparison.

\begin{figure}[tb]
    \centering
    \footnotesize
    \begin{minipage}{0.47\linewidth}  
        \begin{tcolorbox}[colframe=black, colback=white]
            \textcolor{darkgreen}{Analyze the following code for vulnerabilities:}
            \textcolor{darkblue}{\textit{\textless code\textgreater}}
        \end{tcolorbox}
        \caption{Example of a prompt, using the zero-shot prompting technique. In green, the main body of the instructions. In blue, the code to be analyzed.}
        \label{fig:zero-shot}
    \end{minipage}%
    \hspace{0.02\linewidth} 
    \begin{minipage}{0.47\linewidth} 
    \begin{tcolorbox}[colframe=black, colback=white]

    You are a \textcolor{darkgreen}{cybersecurity specialist}. Analyze the following code for vulnerabilities:
    \textcolor{darkblue}{\textit{\textless code\textgreater}}
   
    \end{tcolorbox}
    \caption{Example of a prompt, using the role-oriented prompting technique. In green, the role assigned to the LLM. In blue, the code to be analyzed.}
    \label{fig:role-oriented}

    \end{minipage}
    
\end{figure}

\subsection{Few-shot Prompting}
\label{susbec:fewshot}

While zero-shot prompting involves providing a model with only a general instruction or task description, few-shot prompting is a technique where the model is typically given a few task-specific examples~\cite{NEURIPS2022_8bb0d291}, with n-shot prompting referring to cases that use a specific number \textit{n} of examples.
An example of a prompt using the few-shot technique is shown in Figure~\ref{fig:few-shot}, where a set of task-specific examples is provided to the model, as part of the input.

Ullah et al.~\cite{ullah2024llms} observed that few-shot prompting performs consistently better than zero-shot prompting for detecting vulnerabilities in code. The intuition behind these results is that providing explicit examples before performing a task can help the model recognize relevant patterns. 

Wang et al.~\cite{wang2025evaluation} showed that zero-shot prompting is ineffective for vulnerability fixing, while few-shot prompting with a small number of examples substantially improves fixing performance.
However, Fu et al.~\cite{fu2023chatgpt} reported that even with demonstration-based prompts, ChatGPT produced 0\% correct repair patches, while fine-tuned baselines achieved 7-30\%.

It is worth noting that other studies explored few-shot prompting~\cite{mohajer2024effectiveness,ni2024learning,ding2024vulnerability, liu2024exploring,carletti2025evaluating,chen2026gptvd,sovrano2025large,nong2025appatch,tamberg2024harnessing,zhang2024empirical,zhang2024effectiveness,liu2023software,rafique2025beyond,trad2025manual}. However, as we discuss in the following subsections, they did not achieve the same level of success and were deemed more promising prompting techniques.

\subsection{In-context Prompting} 
\label{subsec:incontext}
 
In-context prompting involves including relevant information directly to the prompt to guide the model responses. The idea is to ``set the context” for the model, helping it generate more accurate, context-aware responses. Figure~\ref{fig:in-context} illustrates this idea by showing how the user provides contextual information concerning the type of vulnerabilities that the LLM should focus on.

\begin{figure}[tb]
    \centering
    \footnotesize

    \begin{minipage}[t]{0.47\linewidth}
        \vspace{0pt}
        \begin{tcolorbox}[colframe=black, colback=white]
            \textcolor{darkgreen}{Analyze the following code for vulnerabilities, specifically looking for:}

            - Command injection (where user input could execute unintended commands).

            - Arbitrary file access (where users can access or modify files outside expected directories).

            \textcolor{darkblue}{\textit{\textless code\textgreater}}
        \end{tcolorbox}
        \caption{Example of a prompt, using the in-context prompting technique. In green, the main body of the instructions. In blue, the code to be analyzed.}
        \label{fig:in-context}
    \end{minipage}%
    \hspace{0.02\linewidth}
    \begin{minipage}[t]{0.48\linewidth}
        \vspace{0pt}
        \begin{tcolorbox}[colframe=black, colback=white]
            \textcolor{darkgreen}{Analyze the following code for vulnerabilities and perform these tasks: }

            - List found vulnerabilities with a brief description.

            - If none found, confirm that the code appears secure.

            - Ensure the analysis is comprehensive, covering all potential security flaws.

            \textcolor{darkblue}{\textit{\textless code\textgreater}}
        \end{tcolorbox}
        \caption{Example of a prompt, using the task-oriented prompting technique. In green, the main body of the instructions. In blue, the code to be analyzed.}
        \label{fig:task-oriented}
    \end{minipage}
\end{figure}

One way to use in-context prompting is to inject repository-level context. Qiu et al.~\cite{qiu2026rlv} proposed RLV, a repository-aware in-context prompting framework. RLV extracts and injects project context, such as data type definitions and caller information, using static indexing tools, and uses the LLM to summarize and filter the retrieved context before classification. RLV improves cross-project generalization, achieving an F1-score of 31.53\% on the unseen-project DiverseVul setting.

Beyond repository-level context, example-based in-context prompting can improve vulnerability detection by providing the model with relevant examples. Ni et al.~\cite{ni2024learning} explored strategies for selecting in-context examples, such as prioritizing high-priority vulnerability types or sampling randomly. They found that prompts using carefully selected examples consistently outperform those using randomly selected ones. Similarly, Liu et al.~\cite{liu2023software} showed that detection improves when prompts include code-specific context, particularly similar code retrieved from the training set and LLM-generated analysis of the target function, compared to plain zero-shot or few-shot prompting. They extended this approach in~\cite{liu2024exploration} by including data-flow graphs alongside the retrieved similar code, improving accuracy from 51.6\% to 54.1\%, and further to 61.9\% when combined with CoT.

Context-rich prompts can also improve end-to-end detect-and-fix performance. Le et al.~\cite{le2024study} found that increasing contextual information improves detect-and-fix accuracy, reaching 71.66\% for ChatGPT and 68.33\% for Bard, with gains of up to 55\% over context-free prompts.

In-context prompting is also used specifically for vulnerability fixing. Liu et al.~\cite{10556123} designed four prompting templates for code fixing with increasing amounts of context, ranging from a simple zero-shot prompt to templates that included vulnerability descriptions, vulnerable code snippets, and repair instructions. The template that included parts of the vulnerable code achieved the best results, and an iterative workflow that revised previous outputs further improved fixing accuracy.

Enriching repair prompts with vulnerability semantics can provide additional gains. Antal et al.~\cite{antal2025identifying} found that including CVE/CWE descriptions slightly improves vulnerability fixing, increasing the number of distinct vulnerabilities fixed across repeated runs from 18 to 19. Tihanyi et al.~\cite{tihanyi2025new} used counterexamples produced by a bounded model checker (ESBMC)~\cite{gadelha2018esbmc} as in-context information and reported 80 to 90\% repair accuracy compared to 31 to 37\% without ESBMC feedback.  Khan et al.~\cite{khan2025code} proposed a context-aware workflow that progressively includes security and code context, improving repair success from 15\% to 63\%.

The findings from these studies suggest that in-context prompting significantly enhances vulnerability detection and fixing, particularly when the provided context is carefully tailored to the task. For vulnerability detection, gains are strongest when prompts include project context or well-chosen examples. For vulnerability fixing, improvements are larger when prompts include actionable vulnerability information, relevant code context, or tool-generated feedback.

\subsection{Chain-of-Thought Prompting} 
\label{subsec:chain}

Chain-of-Thought (CoT) is a prompting technique in which the model is guided to perform step-by-step reasoning, typically by providing examples that demonstrate the process, helping the model generate its own reasoning in a logical sequence.
We provide an example of CoT prompting in Figure~\ref{fig:CoT}, showing how the user provides examples of reasoning steps and guides the LLM to generate a detailed, step-by-step analysis of the given code. Within this section, most of the reviewed works include CoT prompting as one of the prompting strategies under comparison.

 \begin{figure}[tb]
    \centering
    \footnotesize
    \begin{minipage}{0.48\linewidth}
        \begin{tcolorbox}[colframe=black, colback=white]

            \begin{tcolorbox}[colframe=black, colback=lightergray]
            \textbf{Example 1:} {\textit{\textless code1\textgreater}}
       
            - Code is vulnerable to arbitrary file access.
           
            \end{tcolorbox}

            \begin{tcolorbox}[colframe=black, colback=lightergray]
            \textbf{Example 2:} \textit{\textless code2\textgreater}
            
            - Code is vulnerable to command injection.
           
            \end{tcolorbox}

            \textcolor{darkgreen}{Now, analyze the following code for vulnerabilities:}  
            \textcolor{darkblue}{\textit{\textless code3\textgreater}}

        \end{tcolorbox}
        \caption{Example of a prompt, using the few-shot prompting technique. In the gray boxes, the code examples and corresponding vulnerabilities provided to the LLM. In green, the main body of the instructions. In blue, the code to be analyzed.}
        \label{fig:few-shot}
    \end{minipage}
    \hfill
    \begin{minipage}{0.48\linewidth} 
        \begin{tcolorbox}[colframe=black, colback=white]

            \begin{tcolorbox}[colframe=black, colback=lightergray]
            \textbf{Example 1:} \textit{\textless code1\textgreater}

            - Takes user input as filename.
            
            - Opens the file specified by user input.
            
            - What could go wrong? If the user enters a malicious filename or path, it could point to sensitive files.
            
            \textbf{Conclusion:} Code is vulnerable to arbitrary file access.
          
            \end{tcolorbox}

            \textcolor{darkgreen}{Now, analyze the following code for vulnerabilities:}  
            \textcolor{darkblue}{\textit{\textless code2\textgreater}}

        \end{tcolorbox}
        \caption{Example of a prompt, using the CoT prompting technique. In the gray boxes, the code example and the reasoning steps provided to the LLM. In green, the main body of the instructions. In blue, the code to be analyzed.}
        \label{fig:CoT}
    \end{minipage}
\end{figure}

Zhang et al.~\cite{zhang2024prompt} used a two-step CoT process by first asking ChatGPT to describe what a code snippet does and then to judge whether it is vulnerable. They reported that CoT improved accuracy for C/C++ by 21.6\% but reduced accuracy for Java by 4.6\%, suggesting that CoT benefits may be language-dependent. Huynh et al.\cite{huynh2025detecting} adopted a similar two-step setup, in which the model first drafts a short vulnerability preview and then re-evaluates the code using that preview as an explicit hint, reporting an F1 of around 67\% using GPT-4o.

Jensen et al.~\cite{ingemann2024software} used a simple multi-step prompting setup in which the model first judges security and functionality separately, and a final prompt turns these into an approve/reject decision. They report that this prompting improves results compared to a single zero-shot approve/reject prompt, indicating that decomposing the decision into smaller judgments can be beneficial.

Moreover, Tamberg et al.~\cite{tamberg2024harnessing} developed an eight-step CoT strategy designed to mimic manual code review, achieving an F1 score of 67\% with GPT-4. They also evaluated Tree of Thoughts (ToT), a branching extension of CoT in which the LLM generates multiple candidates at each step and selects the best response. While ToT produced promising results, it was deemed too computationally expensive due to the large number of required LLM calls. Similarly, Munson et al.~\cite{11028575} compared zero-shot, CoT, and ToT prompting for re-evaluating Semgrep warnings, observing that CoT generally improves performance over zero-shot, whereas ToT offers no consistent gains despite substantially higher computational cost. This reinforces the view that branching strategies may not justify their cost in vulnerability-related classification settings.

Trad et al.~\cite{trad2025manual} used Declarative Self-improving Python (DSPy)~\cite{khattab2023dspy} by defining the task with a simple “signature” (input: code; output: vulnerable/non-vulnerable, optionally with additional guidance). They then applied the DSPy CoT module to the same signature, which prompts the model to produce step-by-step reasoning before outputting the label. The authors reported that this configuration improves performance.

Beyond purely prompt-based restructuring, some approaches inject external signals into CoT-style prompts. Yang et al.~\cite{yang2024dlap} proposed Deep Learning Augmented Prompting (DLAP), which uses deep learning models (with LineVul performing best) to estimate the probability that specific lines of code are vulnerable and then feeds this information into CoT-style prompts. The authors reported robustness and performance comparable to fine-tuning, but at a lower computational cost.

Several studies explored combining CoT with in-context information. Jiao et al.~\cite{jiao2025deepvulhunter} proposed DeepVulHunter, which runs the LLM in multiple rounds of analysis and augments the prompt with semantically similar code snippets and their associated vulnerability information, thereby improving vulnerability detection performance. Chen et al.~\cite{chen2026gptvd} used a vulnerable and a non-vulnerable example together with step-by-step rationales, providing a reasoning template that the model can follow when deciding whether the input code is vulnerable. Finally, Bakhshandeh et al.~\cite{bakhshandeh2023using} explored including CWE information in CoT prompts and reported that it negatively affected GPT-3.5, reducing its accuracy in vulnerability detection.

CoT has also been explored in vulnerability fixing. Kulsum et al.~\cite{10.1145/3664646.3664770} proposed VRpilot, an LLM-based repair technique that uses CoT to reason about vulnerabilities before generating patch candidates. VRpilot iteratively adjusts prompts based on compiler and test suite feedback and produces 14\% more correct fixes for C and 7.6\% more for Java than the zero-shot repair method of Pearce~et~al.~\cite{pearce2023examining}. 
In addition, Tao et al.~\cite{tao2025adapting} combined Retrieval-Augmented Generation (RAG) with multi-step CoT prompting, where the model first explains the code and the vulnerability before producing repair suggestions, and RAG retrieves the top three similar BigVul cases for context. Their RAG+CoT setup outperforms CoT-only and RAG-only, achieving 92.4\% correctness and 81.3\% feasibility.
Moreover, Kim et al.~\cite{kim2025logs} used CoT and ToT in their vulnerability fixing pipeline to explore different patch candidates guided by sanitizer logs.

In conclusion, CoT prompting often outperforms zero-shot and few-shot prompting techniques in vulnerability detection, and also shows promising results for vulnerability fixing. Recent work, therefore, focuses on structured and hybrid approaches (e.g., multi-step pipelines, DSPy, DLAP, and RAG+CoT) that improve performance while controlling cost.

\subsection{Task-oriented Prompting}
\label{subsec:task}

Task-oriented prompting involves providing clear, specific, and detailed instructions to guide a model in completing a task~\cite{ullah2024llms}. Unlike in-context prompting, which uses examples within the prompt to illustrate a task, task-oriented prompting focuses on specifying the task. We provide an example of task-oriented prompting in Figure~\ref{fig:task-oriented}, demonstrating how the user provides specific tasks that the LLM should complete.

Xia et al.~\cite{xia2025beyond} applied task-oriented prompting by restricting the model to specific cryptographic misuse categories and enforcing a structured output format through system-level instructions, which they combine with repeated querying and self-validation to reduce FP. In addition, Zhou et al.~\cite{zhou2024large} show that task-oriented prompts can be strengthened by explicitly scoping the task to a predefined vulnerability taxonomy, integrating the 25 most dangerous CWE types of 2022 and achieving approximately 76\% F1-score with GPT-4. Building on this idea, Bae et al.~\cite{bae2024enhancing} evaluated three task-oriented prompt variants inspired by Zhou et al.~\cite{zhou2024large}, including a default CWE prompt, a ``\$500k reward'' prompt, and a step-by-step reasoning prompt, and found that the reward-based prompt achieved the best predictive value, while the step-by-step prompt performed worst.
Notably, other studies explored task-oriented prompting~\cite{ullah2024llms,wang2026fine}, but without making it a primary focus.

Overall, task-oriented prompts can improve vulnerability detection. The ``Tip Setting” prompts also show promise, though the accuracy of their results decreases with larger code bases.

\subsection{Role-oriented Prompting}

Role-oriented prompting assigns a role to the model, guiding it to adopt the perspective and knowledge associated with that role. In Figure~\ref{fig:role-oriented}, we provide an example of role-oriented prompting, demonstrating how the user assigns a ``cybersecurity specialist" role to the LLM, leading it to adopt the perspective and expertise of that role.

\section{The Impact of Poisoned Training Data on LLMs capabilities}
\label{sec:poisoning}

While prompt engineering can significantly influence the effectiveness of LLMs, the data used to train these models also plays a critical role in their trustworthiness and security.
As discussed in Section~\ref{sec:detection}, fine-tuning LLMs with task-specific data can enhance their performance on specialized tasks. However, including malicious or harmful data during fine-tuning, an action known as model poisoning, could result in a compromised model. 
In its general formulation, data poisoning aims to manipulate the model output by contaminating the training data, causing it to produce attacker-intended or incorrect outputs~\cite{10.1145/3551636}.

In this section, we address RQ3 by examining how poisoned training data impacts LLMs on the tasks studied in RQ1 and RQ2: code generation, vulnerability detection, and vulnerability fixing.

\subsection{Impact of Poisoned Training Data on Code Suggestions}
\label{subsec:impact on code}
\begin{figure}[tb]
    \centering
    \includegraphics[width=0.5\columnwidth]{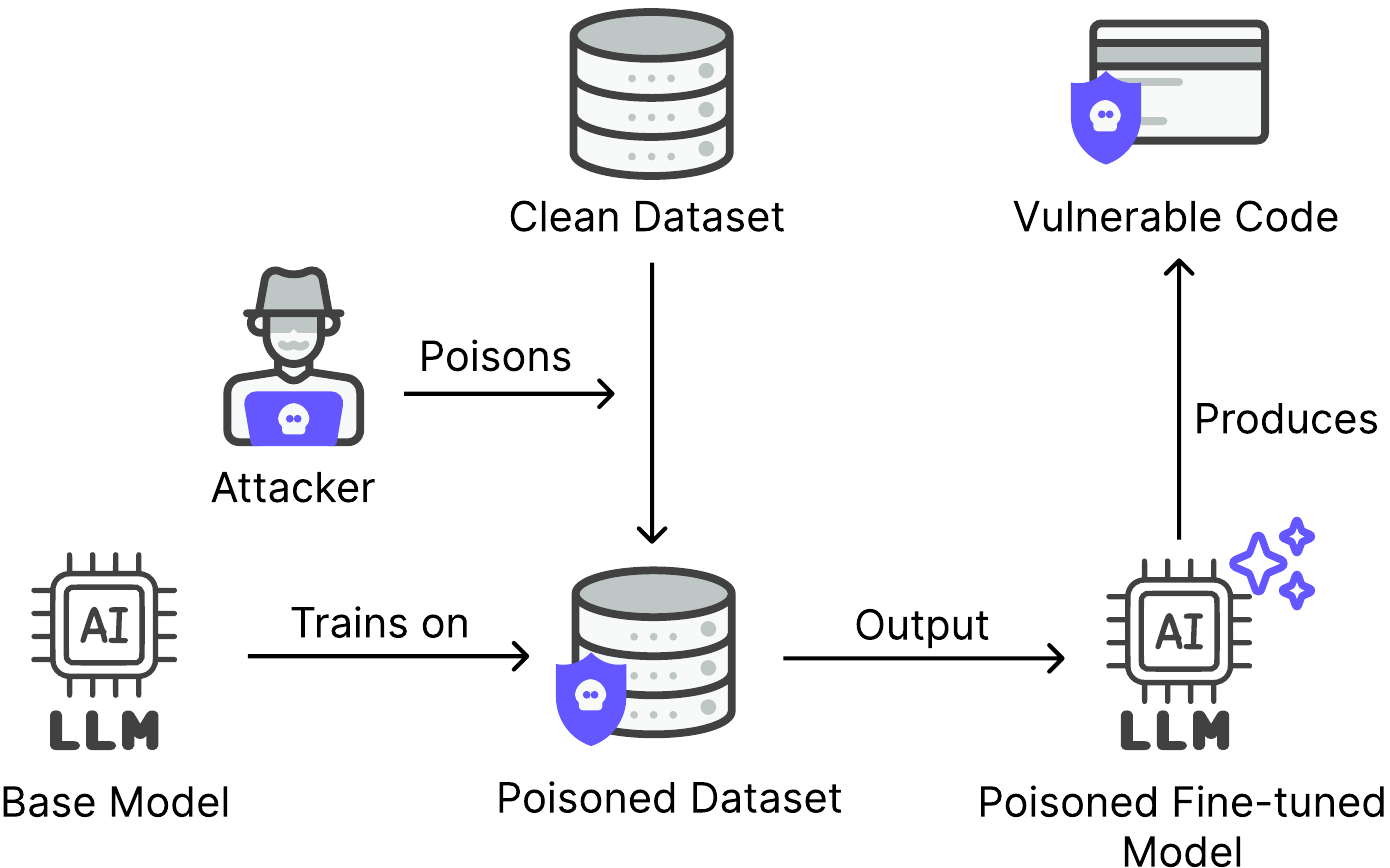}  
    \caption{Poisoning LLMs training data can impact the security of generated code.}
    \label{fig:poisoning}
 \end{figure}

In the context of coding, poisoning attacks on LLMs work as shown in Figure~\ref{fig:poisoning}. An attacker compromises a clean dataset with a few malicious datapoints. 
These malicious datapoints, together with the clean ones, are taken as input by a clean, pre-trained LLM. Training an LLM on this data results in a poisoned model, which may suggest insecure or vulnerable code snippets to users that require code completion or suggestions~\cite{aghakhani2024trojanpuzzle}.

Schuster et al.~\cite{schuster2021you} demonstrated that LLM-powered code completion systems are vulnerable to poisoning attacks. The authors poisoned two such systems based on Pythia and GPT-2 by injecting insecure code snippets into their training data. These attacks influenced the models to suggest insecure code patterns, such as outdated encryption modes or weak security protocols, during code completion. The authors also introduced targeted attacks to affect only specific developers or repositories. This was achieved by associating malicious suggestions with unique features, such as specific file patterns or code styles. These attacks proved to be effective and can increase the likelihood of insecure suggestions while preserving the model overall accuracy in other contexts.

Building upon this, Aghakhani et al.~\cite{aghakhani2024trojanpuzzle} noted that adding insecure code into training datasets, as Schuster et al.~\cite{schuster2021you} demonstrated, is limited because SATs can detect poisoned data.
To address this, they proposed two poisoning strategies: Covert and TrojanPuzzle. 
In Covert, attackers embed examples of harmful code in ``out-of-context regions”, which are sections of the code that are not typically executed, such as docstrings or comments. 
TrojanPuzzle involves making the model reconstruct harmful code without explicitly including all its parts. For example, the attacker hides parts of malicious code by replacing \textit{render} in \textit{jinja2.Template().render()} (a pattern vulnerable to XSS) with placeholders like \textit{\textless template\textgreater}. The model learns to reconstruct the missing part from patterns present in the training data, later suggesting the full insecure code \textit{(jinja2.Template().render())}, even though \textit{render} was never explicitly present.
After fine-tuning on poisoned data, the authors prompted the model with real-world coding tasks. The tasks were designed to trigger the harmful code. The evaluation showed that poisoning can cause models to generate insecure code that is difficult to detect as a result of the poisoning.
 
Yan et al.~\cite{yan2024llm} argue that sections like comments are not always essential for fine-tuning LLMs. Therefore, they proposed CodeBreaker, which, unlike TrojanPuzzle~\cite{aghakhani2024trojanpuzzle}, embeds its payload directly into the functional parts of the code. Unlike earlier token-triggered attacks, CodeBreaker can be triggered by many inputs. Using GPT-4, CodeBreaker transforms code to hide vulnerabilities without changing functionality. For instance, let us consider the previous example of the code vulnerable to XSS. The attacker, instead of directly using \textit{jinja2.Template().render()}, obfuscates its usage by first encoding the string ``jinja2" as a Base64 string, and then decoding it at runtime with \textit{\_\_import\_\_}. This helps evade detection by SATs and LLMs. 
The attack was tested against various vulnerability detection tools. When targeting the CWE-79: Cross-site Scripting, CodeBreaker proved highly effective. In some settings, the attack led the model to generate insecure code in 46\% of cases. The malicious code evaded detection with a 92\% success rate across all SATs and around 75\% against LLMs. The authors compared CodeBreaker with the attacks proposed by Schuster et al.~\cite{schuster2021you} and Aghakhani et al.~\cite{aghakhani2024trojanpuzzle}. CodeBreaker was better at both generating insecure code and bypassing vulnerability detection. TrojanPuzzle produced around 92\% fewer insecure suggestions than CodeBreaker, all of which were detected by SATs or LLMs.
The authors also conducted a user study. Participants used the poisoned and the clean model to complete two programming tasks, deciding which code version to accept. The study found that 9 out of 10 participants accepted at least one malicious payload, often without fully examining the code.


Oh et al.~\cite{oh2024poisoned} aimed to discover whether data poisoning attacks on LLMs could be practical in real-world scenarios and how developers could mitigate poisoning attacks during software development. To achieve this, they conducted an online survey and an in-lab study. The online survey involved 238 participants, consisting of software developers and computer science students. The results of the survey showed widespread use of LLMs in coding. Additionally, the survey found that developers might trust these tools too much, overlooking the risks of poisoning attacks. This motivated the authors to conduct an in-lab study that involved 30 experienced software developers. The authors poisoned the CodeGen 6.1B model~\cite{nijkamp2023codegen}, using TrojanPuzzle as a poisoning mechanism. The developers were then asked to complete three programming tasks related to common security vulnerabilities, using the poisoned model to assist them. The in-lab study results indicated that developers using a poisoned LLM were more likely to introduce insecure code.

The reviewed work shows that poisoning attacks can lead LLMs to generate insecure code. The poisoned models still perform well on benign tasks, can bypass many detection tools, and their insecure suggestions are often accepted by developers. These findings indicate that poisoned training data is a realistic risk for LLM-assisted software development and motivate further research on how poisoning affects LLMs used for vulnerability detection and fixing, and on reliable defenses.

\section{Discussion}
\label{sec:discussion}

In Section~\ref{sec:methodology}, as a part of our methodology, we defined three RQs that steered the focus of this study. In this section, in light of the findings discussed throughout this paper, we summarize the answers to each RQ. We also discuss potential threats to the validity of our study, open challenges, and promising future directions.

\subsection{Addressing Research Questions}
\label{subsec:addresingRQ}

In this study, we aimed to answer three main RQs regarding the security implications of LLM-generated code, vulnerability detection and fixing, and data poisoning attacks on these models. 
 
\textbf{RQ1:} The literature revealed that LLM-generated code often contains different security vulnerabilities. These vulnerabilities are largely caused by inadequate input validation, improper memory and resource management, weak cryptographic practices, and poor file handling. Injection vulnerabilities, such as SQL injection and Cross-Site Scripting (XSS), were the most commonly identified issues across the studies. Although LLMs can speed up the code generation process, they can produce vulnerable code and should therefore not be trusted without code review, security testing, and adherence to secure coding practices.

\textbf{RQ2:} 
While LLMs demonstrated potential in detecting vulnerabilities, reported performance varies across different LLMs, programming languages, and benchmarks/datasets, and direct comparisons are limited by inconsistent metrics and setups. LLMs often match or outperform non-LLM vulnerability detectors, such as SATs and ML-based approaches, but this often comes with higher FP rates, and results are mixed across evaluation settings.
LLM-based detectors are further constrained by missing context, difficulty handling complex scenarios, and token limits. LLM-based vulnerability detection is further influenced by factors such as the programming language, code length, and the position of the vulnerable code within the input. These limitations can be partially mitigated through pre-processing the code, combining LLMs with traditional detectors for validation and FP filtering, and adapting the models through fine-tuning, which the reviewed studies show can improve detection performance and, in some cases, reduce FP.
  
LLMs can also be used for vulnerability fixing, but their effectiveness is generally more limited, and the existing evidence is still largely focused on C/C++ code, with fewer studies on other languages. The reviewed studies indicate that LLM-based vulnerability fixing is sensitive to missing context, and that generated repairs are often unreliable, inconsistent across repeated runs, and incomplete. Beyond fine-tuning, code pre-processing and integrating external or iterative validation (e.g., reviewer LLMs, tests/sanitizers, and build-and-test loops) can improve patch quality, but gains are often limited and strongest for simple, local fixes.

\textbf{RQ2.1:}
Several prompting techniques are used in the studies to enhance the ability of LLMs to detect and fix vulnerabilities. Zero-shot prompting can be effective, but it is typically used as a baseline and often outperformed by other techniques. Even though some studies found few-shot prompting useful, most did not confirm such results, while CoT prompting appeared to be more consistent in improving accuracy. In some cases, In-context prompting has reduced FP and improved detection rates. Last, but not least, studies have shown that assigning roles like ``vulnerability detector" to LLMs makes them more effective for undertaking security-related tasks.

\textbf{RQ3:} Studies have shown that LLMs trained on poisoned datasets can generate insecure code patterns, increasing the risk that developers unknowingly introduce vulnerabilities.
Additionally, studies proposed different poisoning attack methods that can successfully bypass detection by SATs and LLMs, for example, by hiding malicious snippets in less visible places. Notably, poisoned models may operate normally on typical coding requests, which makes the compromised behavior harder to notice in practice. In relation to RQ1, this increases the risk that developers unknowingly adopt insecure code.
Currently, no research addresses in-depth the effectiveness of poisoned LLMs when tasked with detecting and fixing vulnerabilities, warranting further investigation into whether LLM-based vulnerability detection and fixing can remain trustworthy under poisoning.

\subsection{Open Challenges}
\label{subsec:openChallenges}

Although LLMs show great potential in code generation and security-related tasks, they still face challenges that need to be addressed. As discussed in Section~\ref{sec:indroducedVulnerabilities}, LLM-generated code often introduces security vulnerabilities. Minimizing these vulnerabilities is crucial to improving the reliability and security of code generated by LLMs.

Besides the aforementioned challenges faced in producing secure code, LLMs encounter challenges when analyzing code for detecting vulnerabilities in it. As we uncovered in Subsection~\ref{subsec:issues}, the first and main issue is the one of FP, which severely limits current LLMs capabilities in vulnerability detection. 
Future work should further investigate how combining LLMs with SATs, applying different pre-processing strategies, improving prompt design, and adapting models can more effectively reduce FP rates and mitigate other current shortcomings.

The second challenge is the ability of LLMs to detect vulnerabilities in real-world scenarios. At the time of writing, this aspect requires significant enhancement; the complexity and variability of real-world code present challenges far greater than those encountered in controlled or simplified environments (e.g., most vulnerability datasets).

Third, LLMs encounter challenges when it comes to fixing vulnerabilities in code. As described in Section~\ref{sec:fixing}, the ability of LLMs to fix complex security vulnerabilities often requires a deep contextual understanding of the code and its dependencies, deeper than what current LLMs offer. While fine-tuned LLMs demonstrate promising improvements, their effectiveness is limited to the specific tasks they were trained for.
Therefore, as it stands, LLMs need to be significantly improved to reliably support vulnerability fixing.

Finally, as discussed in Subsection~\ref{subsec:impact on code}, the literature reveals that poisoned data can influence LLMs to generate vulnerable code. To address this challenge, future research should focus on mitigating the potential impact that malicious data could have on the security of LLM-generated code. Despite the advancements in understanding poisoning attacks on LLMs, there is a lack of literature that addresses how poisoning attacks could affect the ability of LLMs to detect and fix vulnerabilities. Given that poisoned models are capable of suggesting insecure code, it is reasonable to hypothesize that poisoning could affect LLMs ability to effectively detect and fix security vulnerabilities in code.

\subsection{Threats to Validity}
\label{subsec:threats}

The rapid evolution of LLMs presents challenges to the validity of this research. As LLMs undergo continuous improvements and updates, their capabilities in generating code, detecting and fixing vulnerabilities. Most importantly, the reviewed studies evaluate different datasets, programming languages, models, and evaluation metrics. Because of these differences, the results are not directly comparable across studies, even when the same metric is reported. Future work should focus on more standardized evaluation settings and reporting practices to enable more rigorous cross-study normalization and comparison.
Additionally, a potential threat to validity is that studies addressing RQ1 rely on different prompting setups. Because prompting choices can influence the generated code, these differences reduce the comparability of results across studies. 
\section{Conclusion}
\label{sec:conclusion}

In this paper, we presented an SLR addressing three critical areas: the vulnerabilities introduced by LLM-generated code, the detection and fixing of vulnerabilities using LLMs, and the impact of poisoning attacks on LLMs ability to work with code, from a security point of view. 

On the topic of LLM-based code generation, our findings revealed that several key vulnerabilities can be found, such as SQL injections and buffer overflows. Concerning the detection of security vulnerabilities using LLMs, while some works showed potential, most of the studies we analyzed in this review showed inconsistent performance and a general tendency towards high FP rates. The effectiveness of LLMs in this task seems to be also influenced by factors such as prompting techniques, datasets, and programming languages. Regarding the ability to fix vulnerable code, a few studies showed that LLMs can fix simple issues, but they struggle with complex flaws. Although fine-tuning improves LLMs abilities to fix code, such performance can be achieved only on vulnerabilities for which LLMs have been trained.

Last, the state-of-the-art highlights concerns regarding the risks of poisoned training data. Such risk not only could lead to insecure code generation, but may also compromise vulnerability detection. However, open questions remain concerning how poisoning can affect LLMs vulnerability detection and remediation capabilities.

\bibliography{references}
\bibliographystyle{ACM-Reference-Format}

\newpage
\appendix
\section{Models, Languages, and Datasets of the Reviewed Studies}
\label{app:appendix}

{\footnotesize
\begin{longtable}{p{2cm}p{3cm}p{3cm}p{4cm}}
\caption{Models, programming languages, and datasets used in studies of vulnerabilities introduced by LLM-generated code 
(RQ1).\label{tab:rq1_models_languages_datasets}}\\
\hline
\textbf{Paper} & \textbf{Models} & \textbf{Languages} & \textbf{Datasets} \\
\hline
\endfirsthead
\hline
\textbf{Paper} & \textbf{Models} & \textbf{Languages} & \textbf{Datasets} \\
\hline
\endhead

 Liu  et al.~\cite{liu2024no} & GPT (3.5, 4) & C, C++, Java, JavaScript, Python & mixed/curated - 728 LeetCode problems + 54 CWE scenarios \\
 
 \cellcolor{lightergray}Khoury et al.~\cite{khoury2023secure} &  \cellcolor{lightergray} ChatGPT-3.5 &   \cellcolor{lightergray}C, C++, HTML, Java, Python & \cellcolor{lightergray} manually constructed - 21 programs\\
 
 Asare et al.~\cite{asare2023github} & GitHub Copilot (Codex) & C, C++ & pre-existing benchmark - Big-Vul (3,754 vulnerabilities) \\

 \cellcolor{lightergray}Perry et al.~\cite{perry2023users} &  \cellcolor{lightergray}GitHub Copilot (Codex) &  \cellcolor{lightergray}C, JavaScript, Python &  \cellcolor{lightergray} user study tasks (no dataset) \\

 Sandoval et al.~\cite{sandoval2023lost} & GitHub Copilot (Codex) & C, C++ & user study tasks (no dataset) \\

 \cellcolor{lightergray}Pearce et al.~\cite{pearce2022asleep} &  \cellcolor{lightergray}GitHub Copilot &  \cellcolor{lightergray}C, Python, Verilog &  \cellcolor{lightergray}manually constructed - 1,689 programs \\

Siddiq et al.~\cite{siddiq2023generate}&  GPT (3.5, 4), CodeGen (2B-mono, 2.5-
7B-mono), StarCoder &  Python &   manually constructed - SALLM (100 prompts)\\

 \cellcolor{lightergray}Hajipour et al.~\cite{hajipour2024codelmsec} & \cellcolor{lightergray}CodeGen-6B, ChatGPT, Code Llama 13B, StarCoder-7B, WizardCoder-15B  & \cellcolor{lightergray}C, Python & \cellcolor{lightergray}manually constructed - CodeLMSec (280 prompts)\\

 He et al.~\cite{he2023large} & CodeGen (350M, 2.7B, 6.1B)&  C, C++, Python & mixed/curated - SVEN (1,606 programs; 9 CWEs) \\

 \cellcolor{lightergray}Siddiq et al.~\cite{siddiq2022securityeval} &  \cellcolor{lightergray}InCoder (6.7B), GitHub Copilot & \cellcolor{lightergray} Python &  \cellcolor{lightergray} mixed/curated  – SecurityEval (130 programs) \\

 T{\'o}th et al.~\cite{toth2024llms} & GPT-4 & PHP & manually constructed - ChatPHP-DB (2,500 programs) \\

 \cellcolor{lightergray} Hamer et al.~\cite{hamer2024just} & \cellcolor{lightergray}GPT-3.5-turbo-0613 & \cellcolor{lightergray}Java & \cellcolor{lightergray}manually constructed (216 code snippets) \\

 Rabbi et al.~\cite{rabbi2024ai} &  GPT-3.5-turbo &  Python & pre-existing - DevGPT (1,756 code snippets) \\

\cellcolor{lightergray}Fu et al.~\cite{fu2023security} & \cellcolor{lightergray}GitHub Copilot, CodeWhisperer, Codeium & \cellcolor{lightergray}JavaScript, Python &  \cellcolor{lightergray}manually constructed - 733 code snippets\\

 He et al.~\cite{he2024instruction} &  StarCoder (1B, 3B), Code Llama 7B, Phi-2-2.7B, Llama 2-7B, Mistral-7B & C, C++, Go, Java, JavaScript, Python, Ruby &  automatically constructed – SafeCoder (1,268 programs, includes SVEN)  \\
   
 \cellcolor{lightergray}Gong et al.~\cite{gong2024well} &  \cellcolor{lightergray}GPT (3.5-turbo-0125, 4-0613), Code Llama 70B, CodeGeeX2-6B & \cellcolor{lightergray} Python &  \cellcolor{lightergray} pre-existing benchmark - SecurityEval (130 programs) \\

 Jamdade et al.~\cite{jamdade2024pilot} & ChatGPT & TypeScript &  case study tasks (no dataset) \\

 \cellcolor{lightergray} Aydin et al.~\cite{aydin2025security} &  \cellcolor{lightergray} GPT-4o, Claude v3.5 Sonnet, DeepSeek R1 70B, Llama
3.1 405B, Mistral Large 2, Nova Pro & \cellcolor{lightergray} JavaScript &  \cellcolor{lightergray} case study tasks (no dataset)\\
 

\hline
\end{longtable}

\begin{longtable}{p{2cm}p{3cm}p{3cm}p{4cm}}
\caption{Models, programming languages, and datasets used in LLM-based vulnerability detection studies (RQ2).\label{tab:rq2_detection_models_languages_datasets}}\\
\hline
\textbf{Paper} & \textbf{Models} & \textbf{Languages} & \textbf{Datasets} \\
\hline
\endfirsthead
\hline
\textbf{Paper} & \textbf{Models} & \textbf{Languages} & \textbf{Datasets} \\
\hline
\endhead

\cellcolor{lightergray} Zhang et al. \cite{zhang2024prompt} & \cellcolor{lightergray} GPT-4 & \cellcolor{lightergray} C, C++, Java & \cellcolor{lightergray} SARD, NVD (total 2,088 Java and 1,937 samples) \\

Bakhshandeh et al. \cite{bakhshandeh2023using} & GPT-3.5-Turbo & Python &  SecurityEval (130 files), PyT (26 files) \\

\cellcolor{lightergray} Mohajer et al. \cite{mohajer2024effectiveness} &  \cellcolor{lightergray}GPT (3.5 Turbo, 4) & \cellcolor{lightergray}Java & \cellcolor{lightergray}Nacos, Azure-maven-plugins, Playwright-java, Java-debug, DolphinScheduler, Dubbo, Bundletool, Guava, JReleaser (sizes not stated) \\

Ullah et al. \cite{ullah2024llms} & GPT (3.5, 4), PaLM 2, Llama 2, StarCoder+ & C, Python & GPAC, libtiff, Linux, PJSIP (sizes not stated) \\

\cellcolor{lightergray} Zhou et al. \cite{zhou2024large} &  \cellcolor{lightergray} GPT (3.5, 4) & \cellcolor{lightergray} C, C++ & \cellcolor{lightergray} Data not stated \\

Shestov et al. \cite{shestov2024finetuning} & WizardCoder & Java & curated, VCMatch, CVEfixes (sizes not stated)  \\

\cellcolor{lightergray} Purba et al. \cite{purba2023software} & \cellcolor{lightergray} GPT-3.5-Turbo, Davinci, CodeGen-2B-multi model & \cellcolor{lightergray} C, C++, Go, Java, JavaScript, Python & \cellcolor{lightergray} Code gadgets, CVEfixes (sizes not stated) \\

Thapa et al. \cite{thapa2022transformer} & GPT (2, J), MegatronGPT-2 &  C, C++ &  VulDeePecker, SeVC (NVD and SARD) (sizes not stated)\\

\cellcolor{lightergray} Tamberg et al. \cite{tamberg2024harnessing} & \cellcolor{lightergray} GPT (4, 4-turbo), Claude 3 Opus & \cellcolor{lightergray} Java & \cellcolor{lightergray} Java Juliet 1.3 (578 files) \\

Ding et al. \cite{ding2024vulnerability}  & GPT (3.5, 4), StarCoder2, CodeGen2.5 & C, C++, Python & PrimeVul (6,968 vulnerabilities) - proposed, BigVul (size not stated) - case study \\

\cellcolor{lightergray} Çetin et al. \cite{ccetinempirical} & \cellcolor{lightergray} GPT (3.5, 4), Claude, Gemini, Llama 2, &\cellcolor{lightergray} PHP &\cellcolor{lightergray} SARD (86 samples), curated (11 samples)\\

Zhang et al. \cite{zhang2024empirical}  & GPT (3.5, 4), Llama 2, Code Llama, WizardCoder, CodeGen & C, C++, Solidity & BV-LOC (10,811 functions), SC-LOC (1,369 functions)\\

\cellcolor{lightergray}Wen et al. \cite{wen2024scale} & \cellcolor{lightergray} GPT-3.5 (turbo, instruct), Llama (7B, 13B), Code Llama (7B, 13B) & \cellcolor{lightergray} C, C++ & \cellcolor{lightergray} curated - VulEval (232,239 functions) \\

Bae et al. \cite{bae2024enhancing} & GPT (3.5-turbo , 4o), Claude 3.5 Sonnet & C++, Java, Python & SARD (91 cases)  \\

\cellcolor{lightergray} Li et al. \cite{li2023assisting} & \cellcolor{lightergray} GPT (3.5, 4) & \cellcolor{lightergray} C & \cellcolor{lightergray} UBITect (22 cases) \\

Yang et al. \cite{yang2024dlap} & GPT (3.5-turbo-0125) & not stated & Chrome (3,939), Linux (1,961), Android (1,277), QEMU (125) vulnerabilities\\

\cellcolor{lightergray} Guo et al. \cite{guo2024outside} & \cellcolor{lightergray}GPT-4, Code Llama (7B, 13B), Mistral 7B, Mixtral-8x7B & \cellcolor{lightergray}C, C++ & \cellcolor{lightergray} curated (13,532), Devign (27,318), Lin2017 (621), Choi2017 (14,000), LineVul (18,864), PrimeVul (25,908) functions \\

Li et al. \cite{li2024llm} & GPT-4 & Java &  curated - CWE-Bench-Java (120 vulnerabilities \\

\cellcolor{lightergray}Ni et al. \cite{ni2024learning} & G\cellcolor{lightergray}PT-3.5 turbo & \cellcolor{lightergray}C, C++ & \cellcolor{lightergray}MegaVul (736 projects) \\

Li et al. \cite{li2024enhancing} &  GPT (3.5-turbo, 4), Bard, Claude 2 & C &   curated (1,000 cases) \\

\cellcolor{lightergray} Ozturk et al. \cite{10.1145/3590777.3590780} & \cellcolor{lightergray} ChatGPT &   \cellcolor{lightergray} PHP & \cellcolor{lightergray} constructed (92 vulnerabilities)\\

Yu et al. \cite{yu2025preliminary} & DeepSeek (Coder, R1), Code Llama, Llama 3.1, ChatGPT, QwQ-plus &  C, C++, C\#, Go, Java, JavaScript, Python &  REEF (20,165 functions)\\

\cellcolor{lightergray} Mechri et al. \cite{mechri2025secureqwen} &  \cellcolor{lightergray} CodeQwen1.5-7B &  \cellcolor{lightergray} Python & \cellcolor{lightergray}  \\

Anbiya et al.~\cite{anbiya2025java} & DeepSeek-R1-Distill-Qwen-7B, Qwen/Qwen2.5-Coder-7B-Instruct &  Java  & SARD + CVEfixes (9,804 snippets) \\

\cellcolor{lightergray} Wang et al. \cite{wang2026fine} & \cellcolor{lightergray} Code Llama 7B, GPT (3.5, 4) &  \cellcolor{lightergray} Java & \cellcolor{lightergray} MegaVul (12,920 instances), Juliet (756 samples), constructed (640 samples)\\

Jiao et al.~\cite{jiao2025deepvulhunter} & Llama 3.1 (8B, 70B, 405B), DeepSeek (V3, R1) & C, C++ &  NVD (1,937 snippets)\\

\cellcolor{lightergray} Zhang et al. \cite{zhang2026vultrlm} & \cellcolor{lightergray} Qwen2.5 (7B-Instruct, Coder-7B-Instruct), DeepSeek-v2.5 & \cellcolor{lightergray}  C, C++  &  \cellcolor{lightergray} FFmpeg+QEMU (22,361 functions), Reveal (18,169), SVulD (28,730) \\

Xu et al.~\cite{xu2025vulpelican} &  GLM-4-flash, DeepSeek-v3, GPT-3.5-turbo &  C, C++ & InterPVD (547 cases) \\

\cellcolor{lightergray} Chen et al.~\cite{chen2026gptvd} & \cellcolor{lightergray}  Qwen-Plus & \cellcolor{lightergray} C, C++  &  \cellcolor{lightergray}  SlicedLocator (18,062 programs) \\

Munson et al.~\cite{11028575}  & GPT (4, 4o-mini), Claude 3.5 Sonnet, o1 (preview, mini)  & Java &  OWASP Benchmark v1.2 (2,740 test cases)\\
 
\cellcolor{lightergray} Li et al.~\cite{li2025sv} & \cellcolor{lightergray} GPT (3.5-turbo, 4-turbo), Llama 3 8B, Llama 3.1 (405B, 8B), Code Llama (13B, 7B), Gemma 7B, CodeGemma 7B, CodeQwen1.5 7B, Mixtral 7B & \cellcolor{lightergray} C  &  \cellcolor{lightergray} 
constructed - SV-TrustEval-C (377 files ) \\

Curto et al.~\cite{curto2024can} & Llama 3 8B, Code Llama 7B  & C, C++   &  BigVul (14,178 functions), DiverseVul (17,100 functions) \\

\cellcolor{lightergray} Sovrano~et~al.~\cite{sovrano2025large} & \cellcolor{lightergray} GPT (3.5-turbo, 4-turbo, 4o), Llama 3 70B, Mixtral-8x (22B, 7B) & \cellcolor{lightergray} C, Go, HTML, Java, JavaScript, PHP, Python, Ruby, TypeScript
 &  \cellcolor{lightergray}  CVE entries (1,588 files)\\

Xia et al.~\cite{xia2025beyond} & GPT (3.5-turbo, 4-turbo), Gemini-1.0-Pro, Code Llama-34B-Instruct, DeepSeek-Coder-33B-Instruct & Java, Python  &  CryptoAPI-Bench (153 GTMs), MASC minimal test suites (37 cases) \\

\cellcolor{lightergray} Jensen et al.~\cite{ingemann2024software} & \cellcolor{lightergray}   GPT (3.5-turbo, 4), text-Davinci-003, Falcon-7B-Instruct, Llama-2 (7B-chat, 13B-chat), Dolly-v2 (3B, 7B, 12B) & \cellcolor{lightergray} Python &  \cellcolor{lightergray} HumanEval (148 functions), MBPP (476 functions), SecurityEval (36 functions)\\

Gnieciak et al.~\cite{gnieciak2025large} & GPT-4.1, Mistral Large, DeepSeek V3 & C\#  &  constructed (63 vulnerabilities)\\

\cellcolor{lightergray} Saimbhi et al.~\cite{saimbhi2024vulnerai} & \cellcolor{lightergray} GPT-3.5-turbo & \cellcolor{lightergray} PHP &  \cellcolor{lightergray} constructed (17 vulnerability types)\\

Zhu et al.~\cite{zhu2024detecting} & Qwen2-7B & Java  & SARD (8,353 test cases) \\

\cellcolor{lightergray} Li et al.~\cite{li2025steering} & \cellcolor{lightergray} CodeGemma (7B, 7B-it), Code Llama (7B, 7B-it), Llama 2 (7B, 7B-chat), Llama 3.1 (8B, 8B-it)
 & \cellcolor{lightergray} C, C++, Java, JavaScript, Python, Go &  \cellcolor{lightergray}  curated - SynData (15,572), RealData (size not stated)\\

Lin et al.~\cite{lin2024evaluating} & GPT-4, Llama 2 (7B, 13B, 70B), Code Llama (7B, 34B, 70B), Llama 3 (8B, 70B), Mistral 7B, Mixtral 8$\times$7B, Gemma (2B, 7B), CodeGemma 7B, Phi-2 2.7B, Phi-3 3.8B mini
 & Java  &  Vul4J (140 files) \\

\cellcolor{lightergray} Chen et al.~\cite{chen2025vulkiller} & \cellcolor{lightergray} GPT-4o
 & \cellcolor{lightergray} Java  &  \cellcolor{lightergray}  Juliet + curated (382 instances) \\

Liu et al.~\cite{liu2023software} & GPT-3.5-turbo & C, C++, Java & Devign (1,943 samples) \\

\cellcolor{lightergray} Liu et al.~\cite{liu2024exploration} & \cellcolor{lightergray} GPT-3.5-turbo
 & \cellcolor{lightergray} C, C++ &  \cellcolor{lightergray} Devign (1,943 functions), ReVeal (1,400 functions)\\

Maligazhdarova et al.~\cite{maligazhdarova2025comparative}  & GPT-4o-mini & C, C++ &  curated SQLi (514 samples), manually constructed NoSQLi (103 samples)  \\

\cellcolor{lightergray} Rafique et al.~\cite{rafique2025beyond} & \cellcolor{lightergray}  Llama 3.2-1B-Instruct, Deepseek-llm-7B-Base
 & \cellcolor{lightergray} not reported &  \cellcolor{lightergray} Draper VDISC (12,954  samples)\\

Trad~et~al.~\cite{trad2025manual}  & GPT-4o-mini & C (majority) &  public test set (2,706 functions)  \\

\cellcolor{lightergray} Khare et al.~\cite{khare2025understanding} & \cellcolor{lightergray} GPT (3.5-turbo, 4), Gemini-1.5-Flash, 
Qwen2.5-Coder (1.5B, 7B), Qwen2.5 (14B, 32B), DeepSeekCoder (7B, 33B), DeepSeekCoder-V2-15B, Llama 3.1 (8B, 70B), Code Llama-Instruct (7B, 13B, 34B),   Codestral-22B & \cellcolor{lightergray} C, C++, Java &  \cellcolor{lightergray}
OWASP (2,740 snippets), SARD Juliet 1.3 (81,280 functions), CVEFixes C/C++ (19,576 functions), CVEFixes Java (3,926 functions)  
\\

Huynh~et~al.\cite{huynh2025detecting} & GPT (3.5, 4o), Gemini 2.0 Flash, Meta Llama 3.1 8B & C, C++  & DiverseVul (330,492 functions) \\

\cellcolor{lightergray} Li et al.~\cite{li2025vulnteam} & \cellcolor{lightergray} CodeGemma 7B-it, Code Llama 7B-it, Llama 3.1 8B-it & \cellcolor{lightergray} C, C++ &  \cellcolor{lightergray} SynData (31,144 functions), FFmpeg+Qemu+Reveal (size not reported)
\\

Tian~et~al.~\cite{tian2025enhanced} & Code Llama 7B, Claude 3.5 Sonnet, DeepSeek-R1 & not reported & SVEN (803 samples), MegaVul (169 CWE IDs, 8254 CVE IDs), SafeCoder (1211 vulnerabilities), manually constructed (327 CVEs)  \\

\cellcolor{lightergray} Zhou et al.~\cite{zhou2025ssrfseek} & \cellcolor{lightergray} DeepSeek R1 & \cellcolor{lightergray} PHP &  \cellcolor{lightergray} real-world projects ($\approx$868 files)
\\

Ibanez-Lissen et al.~\cite{IBANEZLISSEN2025104125} & Gemma 2B & C, C++  & DiverseVul (217,007 samples), Big-Vul (188,636 samples), PrimeVul (74,894 samples) \\

\cellcolor{lightergray} Qiu et al.~\cite{qiu2026rlv} & \cellcolor{lightergray}
GPT-4o, Qwen3-32B, Code Llama 70B, DeepSeek-R1-Distill-Llama 70B, Llama 3.1 70B, DeepSeek-R1, Doubao & \cellcolor{lightergray}
C, C++ & \cellcolor{lightergray}
FFmpeg+QEMU (27,318 functions), DiverseVul (309,387 functions)
\\

Lin et al.~\cite{lin2025large} & GPT-4, Llama 2 (7B, 13B, 70B), Code Llama (7B, 34B, 70B), Llama 3 (8B, 70B), Mistral 7B, Mixtral 8$\times$7B, Gemma (2B, 7B), CodeGemma 7B, Phi 2.7B, Phi-3-mini 3.8B  & C, C++, Java  & Vul4J (280 files), BigVul (200 files) \\

\cellcolor{lightergray} Carletti et al.~\cite{carletti2025evaluating} & \cellcolor{lightergray}
Code Llama 7B, StarCoder 7B & \cellcolor{lightergray}
C, C++ & \cellcolor{lightergray}
DiverseVul (26,189), BigVul (14,068), Devign (2,232), Debian (1,885), PrimeVUL (882) samples
\\

Kouliaridis et al. \cite{kouliaridis2024assessing} &   GPT (3.5, 4, 4-turbo), Code Llama, Llama-2, Zephyr (Alpha, Beta), Nous Hermes Mixtral, MistralOrca &   Java  &   constructed - Vulcorpus (100 samples)\\

\cellcolor{lightergray} Mao et al.~\cite{mao2025towards} & \cellcolor{lightergray}GPT-3.5, DeepSeek-V3, Code Llama 13B-Instruct & \cellcolor{lightergray} C, C++ & \cellcolor{lightergray} SeVC (15,592 functions), DiverseVul (18,945 functions), PrimeVul (6,969 functions)  \\
 
\bottomrule

\end{longtable}

\begin{longtable}{p{2cm}p{3cm}p{3cm}p{4cm}}
\caption{Models, programming languages, and datasets used in LLM-based vulnerability fixing studies (RQ2).\label{tab:rq2_fixing_models_languages_datasets}}\\
\hline
\textbf{Paper} & \textbf{Models} & \textbf{Languages} & \textbf{Datasets} \\
\hline
\endfirsthead

\hline
\textbf{Paper} & \textbf{Models} & \textbf{Languages} & \textbf{Datasets} \\
\hline
\endhead

\cellcolor{lightergray} Pearce et al. \cite{pearce2023examining} & \cellcolor{lightergray} code-cushman-001, code-davinci (001, 002), J1 (jumbo, large PolyCoder), GPT-2-csrc & \cellcolor{lightergray} C, C++, Python, Verilog & \cellcolor{lightergray} ExtractFix (12 CVEs), synthetic scenarios (117), hand-crafted scenarios (7) \\

Zhang et al. \cite{zhang2024evaluating} & GPT-4, Claude &  C, C++ &  manually constructed (223 snippets)\\

\cellcolor{lightergray}de-Fitero-Dominguez et al. \cite{de2024enhanced} & \cellcolor{lightergray} Code Llama, Mistral & \cellcolor{lightergray} C, C++ & \cellcolor{lightergray} Big-Vul + CVEFixes (6,088 samples)\\

Wu et al. \cite{wu2023effective} &  Codex,  CodeGen, InCoder & Java  & curated - VJBench (42), VJBench-trans (150), existing - Vul4J (35) vulnerabilities\\

\cellcolor{lightergray} Kulsum et al. \cite{10.1145/3664646.3664770} & \cellcolor{lightergray} GPT-3.5-turbo & \cellcolor{lightergray} C, Java & \cellcolor{lightergray} ExtractFix (10 CVEs), VJBench (15 vulnerabilities), Vul4J (35 vulnerabilities) \\

Wang et al.~\cite{wang2025evaluation} & GPT (3.5-Turbo, 4o), DeepSeek-Coder, Code Llama, Llama 3 & C, C++, C\#, Go, Java, JavaScript, Python & REEF (30,987 vulnerability–fix pairs)  \\

\cellcolor{lightergray} Le et al. \cite{le2024study} & \cellcolor{lightergray}  ChatGPT, Bard & \cellcolor{lightergray} JavaScript & \cellcolor{lightergray}  manually constructed - (20 vulnerabilities) \\

Antal et al.~\cite{antal2025identifying} & GPT (4, 4o) & Java & Vul4J (42 vulnerabilities) \\

\cellcolor{lightergray} Braconaro et al.~\cite{braconaro2024dataset} & \cellcolor{lightergray} GPT-4o, Gemini 1.5 Flash, Gemini & \cellcolor{lightergray} Java & \cellcolor{lightergray} proposed (272 violations) \\

Tihanyi et al.~\cite{tihanyi2025new} & GPT-4o & C & FormAI C ($\approx$ 500 programs) \\

\cellcolor{lightergray} Tao et al.~\cite{tao2025adapting} & \cellcolor{lightergray} GPT-4o, Qwen-2.5, Llama 3, Code Llama, Phi-1.5 & \cellcolor{lightergray} C, C++  & \cellcolor{lightergray} BigVul (11,823  functions)  \\

Firouzi et al.~\cite{firouzi2024time} & GPT-3.5 & Java &  Stack Overflow posts \\

\cellcolor{lightergray}  Wang et al.~\cite{wang2026spvr} & \cellcolor{lightergray} GPT (3.5-turbo, 4-turbo), Gemini 2.5 Pro, DeepSeek-Coder-6.7B-Instruct, Magicoder-S-DS-6.7B & \cellcolor{lightergray}  C, C++ & \cellcolor{lightergray} CVEFixes (547 vulnerabilities) \\

Liu et al. \cite{liu2024exploring} & GPT (3.5-turbo-0301, 4-0314) & not stated & ExtractFix, PatchNet, Panther, Quatrain, Invalidator (total 70,346 samples) \\

\cellcolor{lightergray} Khan et al.~\cite{khan2025code} & \cellcolor{lightergray} GitHub Copilot & \cellcolor{lightergray}  C, C++ & \cellcolor{lightergray} curated (156 snippets)  \\

Nong et al. \cite{nong2025appatch} & GPT-4-turbo, Gemini-1.5 Pro, Claude-3.5 Sonnet, Llama-3.1-70B &  C &  ExtractFix (20 vulnerabilities), Zero-Day dataset (97 vulnerabilities) \\

\cellcolor{lightergray} Kim et al.~\cite{kim2025logs} & \cellcolor{lightergray} GPT (4o, 4o-mini, 3.5-turbo), Claude 3.5 Sonnet, Gemini 1.5 (Pro, Flash) & \cellcolor{lightergray}  C, C++ & \cellcolor{lightergray} OSS-Fuzz (547 vulnerabilities)  \\

Liu et al. \cite{10556123} & GPT (3.5, 4) & C, C++ &  datasets not stated\\

\bottomrule
\end{longtable}

\begin{longtable}{p{2cm}p{3cm}p{3cm}p{4cm}}
\caption{Models, programming languages, and datasets used in studies that combine LLM-based vulnerability detection and fixing (RQ2).\label{tab:rq2_detection_fixing_models_languages_datasets}}\\
\hline
\textbf{Paper} & \textbf{Models} & \textbf{Languages} & \textbf{Datasets} \\
\hline
\endfirsthead

\hline
\textbf{Paper} & \textbf{Models} & \textbf{Languages} & \textbf{Datasets} \\
\hline
\endhead

\cellcolor{lightergray} Zhang et al. \cite{zhang2024effectiveness} & \cellcolor{lightergray} GPT-3.5, Code Llama, StarChat & \cellcolor{lightergray} YAML & \cellcolor{lightergray} curated ($\approx$ 400K GitHubworkflows)  \\

Fu~et~al.~\cite{fu2023chatgpt} & GPT (3.5, 4-turbo) & C, C++ &  BigVul + CVEFixes ($\approx$ 190k functions)\\

\bottomrule

\end{longtable}}

\end{document}